\documentclass[showpacs,preprint]{revtex4}
\usepackage{tabularray}
\usepackage{graphicx}
\usepackage{bm}
\usepackage{placeins}
\usepackage{xspace}
\usepackage{bm,amsmath,amssymb,amsfonts,gensymb,bbold}
\usepackage{comment}
\usepackage{hyperref}
\hypersetup{
    colorlinks=true,
    linkcolor=blue,
    citecolor=blue,
    filecolor=blue,      
    urlcolor=blue,
}
\usepackage{nicefrac}
\usepackage{cleveref}

\usepackage{color}

\usepackage[english]{babel}
\usepackage{verbatim}
\usepackage{makecell}

\usepackage{ amssymb }
\begin{document}

\title{Thermodynamics and Shadows of Kerr black holes endowed with a global monopole charge}

\author{Balendra Pratap Singh$^{a}$}\email{balendrap.singh@ddn.upes.ac.in}
\author{Md Sabir Ali$^{b}$}\email{alimd.sabir3@gmail.com}
\affiliation{$^{a}$ Department of Physics, Applied Science Cluster, SOAE, UPES, Energy Acres, Bidholi, Via Prem Nagar,  Dehradun, Uttarakhand, 248007, India}
\affiliation{$^{b}$ Department of Physics, Mahishadal Raj College, West Bengal 721628, India}

\begin{abstract}
In this article, we present the thermodynamic and shadow properties of rotating black holes with global monopole charge. The angular diameter of Sgr A$^{*}$ black hole is 48.7 $\pm$ 7 $\mu as$, 8 kps far away having a mass of $M = 4.0_{-0.6}^{+1.1} \times 10^6 M\odot$ as observed by Event Horizon Telescope and for the M87 black hole, the observed angular diameter is $\theta_d = 42 \pm 3 \mu$as, which is almost $16$ $Mpc$ far away with a mass of $M = (6.5 \pm 0.7) \times 10^9 M_\odot$. The global monopole charge parameter $\alpha$ strongly affects the shape and size of the black hole shadow. We derived all the necessary equations to obtain the angular diameter of the rotating black hole shadow with the effect of the global monopole charge parameter $\alpha$.   For $\alpha$ 
 $\in$ (0, 0.08) with $a$ $\in$ $(0.7 M, 0.99 M)$, the angular diameter of M87 black hole shadow varies from $39$ $\mu as$ to $51$ $\mu as$. The angular diameter of Sgr A$^{*}$ black hole with global monopole charge parameter $\alpha$  $\in$ (0, 0.04) and $a$ $\in$ $(0.7 M, 0.99 M)$, varies from $50$ $\mu as$ to $55$ $\mu as$. For bound values of $\alpha$ and $a$, our results are consistent with the EHT observations.
\end{abstract}

\maketitle

\section{Introduction}
General relativity undoubtedly is a remarkable gravity theory and is extremely successful on both theoretical and observational background to test solar system and beyond. One of the most bizarre predictions of general relativity is the existence of black holes as outcomes of Einstein's field equations. The first black hole solution was theorized by Karl Schwarzschild in 1916 \cite{Schwarzschild:1916uq}, which is a vacuum solution of Einstein's field equations. The Schwarzschild black hole is a static and spherically symmetric spacetime. Likewise,  we have charged black hole solution known as RN black hole which we obtain when general relativity is minimally coupled to Maxwell's electrodynamics \cite{Reissner:1916cle,Nordstrom:1918bv,Weyl:1917rtf,Jeffery:1921rsl}. The rotating counterparts of the Schwarzschild and Reissner-N$\ddot{o}$rdstrom spacetimes are the Kerr \cite{Kerr:1963ud, Newman:1965tw} and Kerr-Newman black holes \cite{Newman:1965my}, respectively. These four solutions are purely the general relativistic solutions of Einstein field equations and collectively known as the vacuum/electro-vacuum solutions. It is argued that a global monopole charge pertaining to a spontaneous breakdown of the global $O(3)$ symmetry to $U(1)$ is produced during the early inflationary epoch of the phase transitions of the universe \cite{Barriola:1989hx}. When we add the global monopole charge as a possible source of the energy-momentum tensor in Einstein's field equations, we have the global monopole charge-inspired Schwarzschild black hole \cite{Dadhich:1997mh} amounting to the breakdown of the vacuum and asymptotic flatness of the Schwarzschild black hole. However, we still have a black hole solution with a horizon corresponding to the study of the effect of the global monopole charge on the thermal and optical properties of black holes.\\
The thermodynamics of the black hole is one of the possible routes to probe the quantum theory of gravity. The pioneering work of Bardeen, Carter, and Hawking, where they formulated in a geometric way the four laws of black hole mechanics, has revolutionized our understanding of the spacetime geometry \cite{Bardeen:1973gs}. The particle creation by black holes and the correspondence of the area to the entropy of the black hole event horizon, then connecting the surface gravity to the temperature, were undoubtedly a few of the path-breaking discoveries in gravitational physics \cite{Hawking:1975vcx, Bekenstein:1973ur}. In our study, we investigate the basic thermodynamic properties, including Hawking temperature, specific heat, and Gibbs free energy of the rotating Kerr black hole in the presence of the global monopole charge. The contribution of a global monopole to the black hole spacetime comes when we have a spontaneous breakdown of a scalar-field triplet corresponding to a global symmetry group $O(3)$.
The detailed discussions of the black hole solution and its rotating version pertaining to a global monopole charge environment are incorporated in \cite{Singha:2025huj}. \\
The current observations from the Event Horizon Telescope (EHT) of the M87 \citep{EventHorizonTelescope:2019dse, EventHorizonTelescope:2019pgp, EventHorizonTelescope:2019ggy,EventHorizonTelescope:2019jan, EventHorizonTelescope:2019ths, EventHorizonTelescope:2019uob}   and Sgr A$^{*}$  \citep{EventHorizonTelescope:2022exc,EventHorizonTelescope:2022urf,EventHorizonTelescope:2022apq,EventHorizonTelescope:2022wok,EventHorizonTelescope:2022wkp, EventHorizonTelescope:2022xqj} black hole open a new gateway to test various theories of gravity, including Einstein's general theory of relativity. Black holes are completely dark objects, still, these astrophysical objects cast shadow. The image of a black hole's shadow appears as a dark spot surrounded by electromagnetic radiation. The dark spot corresponds to the event horizon of the black hole, and the visible part of the  electromagnetic radiation appears as bright photon rings \citep{EventHorizonTelescope:2019dse, EventHorizonTelescope:2022wkp}.  The incoming light towards a black hole may have three possibilities: the incoming radiations with lower angular momentum fall inside the black hole, the second one is that higher angular momentum photons may scatter to the infinity and the third possibility is that the photos with sufficient angular momentum may form unstable circular orbits around the black hole so called photon orbits \citep{Synge:1966okc,Bardeen:1973tla,Luminet:1979nyg,Cunningham:1973tf}. 
These photon orbits are circular for the spherically symmetric non-rotating black holes and distorted for the axially symmetric rotating black holes \cite{Bozza:2010xqn}. This subject, black hole shadow, has been extensively studied by numerous researchers in the past few years \citep{Falcke:1999pj,Vries2000TheAS,Shen:2005cw,Yumoto:2012kz,Atamurotov:2013sca,Abdujabbarov:2015xqa,Cunha:2018acu,Kumar:2018ple,Afrin:2021ggx,Hioki:2009na,Chen:2023wzv,Li:2024abk, Amarilla:2010zq,Amarilla:2011fx,Amarilla:2013sj,Amir:2017slq,Singh:2017vfr,Mizuno:2018lxz,Allahyari:2019jqz,Papnoi:2014aaa,Kumar:2020hgm,Kumar:2020owy,Kumar:2019ohr, Kumar:2017tdw, Singh:2017xle, Kumar:2020yem,  Kumar:2020sag, Kumar:2020ltt, Kumar:2019pjp, Kumar:2024cnh, Vishvakarma:2023csw, Ghosh:2020spb,Guo:2020zmf,Afrin:2021wlj,Vagnozzi:2022moj,Vagnozzi:2019apd,Afrin:2021imp,Gao:2023mjb,Ghosh:2022jfi,  Li:2022eue,Sengo:2022jif,Liu:2024lbi, Umarov:2025btg, Ahmed:2025zdc, Turakhonov:2025ojy, Singh:2023zmy, Singh:2022dqs, KumarWalia:2022aop}. Some researchers extended this study for quantum inspired theories of black holes \citep{Liu:2020ola,Brahma:2020eos,KumarWalia:2022ddq,Islam:2022wck,Afrin:2022ztr, Yang:2022btw}. Many physicist have considered the higher-dimensional spacetime for analyzing the black hole shadow \citep{Papnoi:2014aaa,Singh:2017vfr,Amir:2017slq,Singh:2023ops,Vagnozzi:2019apd,Hou:2021okc,Vagnozzi:2022tba,Banerjee:2022jog,Lemos:2024wwi, Singh:2024rnh}. Inspired by these works, we study the thermodynamical and shadow properties of a rotating black hole with the effect of the global monopole charge parameter. The global monopole charge parameter strongly affects the shape and size of the black hole shadow, and in the presence of this parameter, we find the angular diameter of the M87 and Sgr A$^{*}$ black holes.
\\
The organization of the paper is as follows. In Sec.~\ref{metric}, we introduce the black hole meric and its horizon properties.  In Sect.~\ref{thermodynamics}, we discuss the thermodynamic properties, including the effects of a global monopole charge and the rotation parameter, e.g., the Hawking temperature, heat capacity, and the Gibbs free energy. In Sect.~\ref {geodesics}, we derive all the geodesic equations and find the effective potential of the black hole. In Sect.~\ref{shadow}, we analytically study the black hole shadow properties and their observables, and we estimate the angular diameter of M87 and Sgr A$^{*}$ black hole in Sect.~\ref{Constraints}.  We finally conclude our results in Sect.~\ref{conclusion}.
\section{Black  Hole Metric}
\begin{figure}
    \centering
    \includegraphics[width=0.4\linewidth]{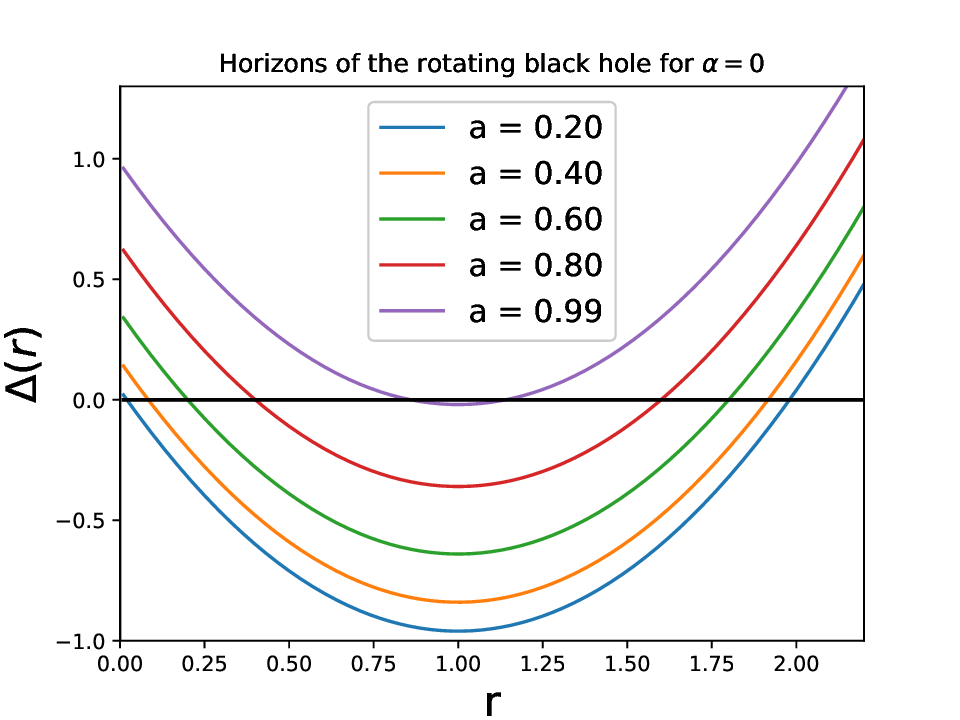}
    \includegraphics[width=0.4\linewidth]{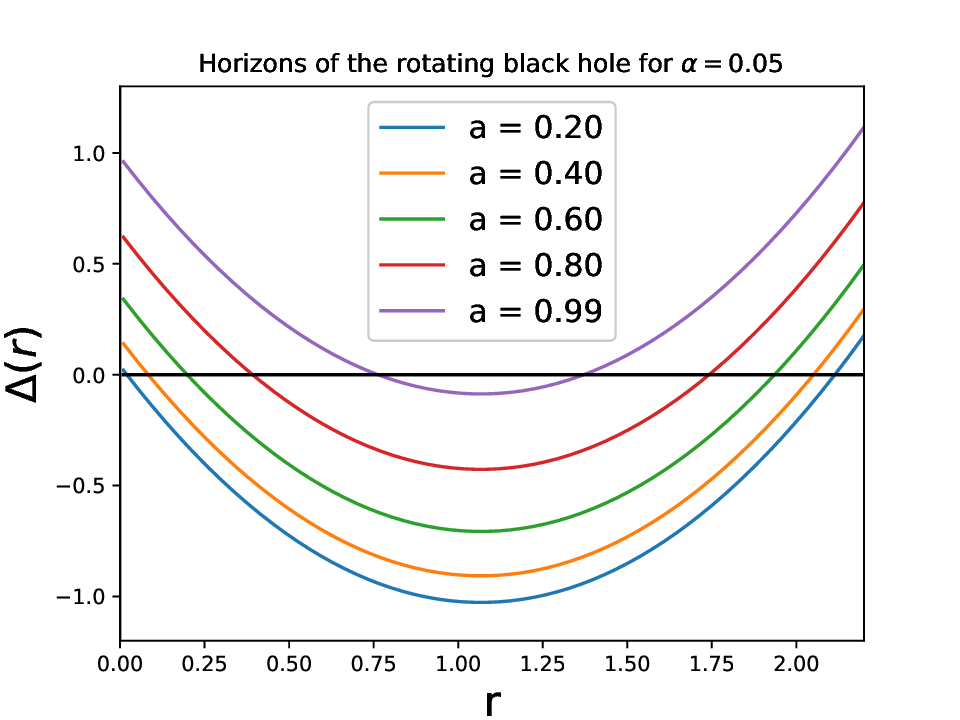}\\
    \includegraphics[width=0.4\linewidth]{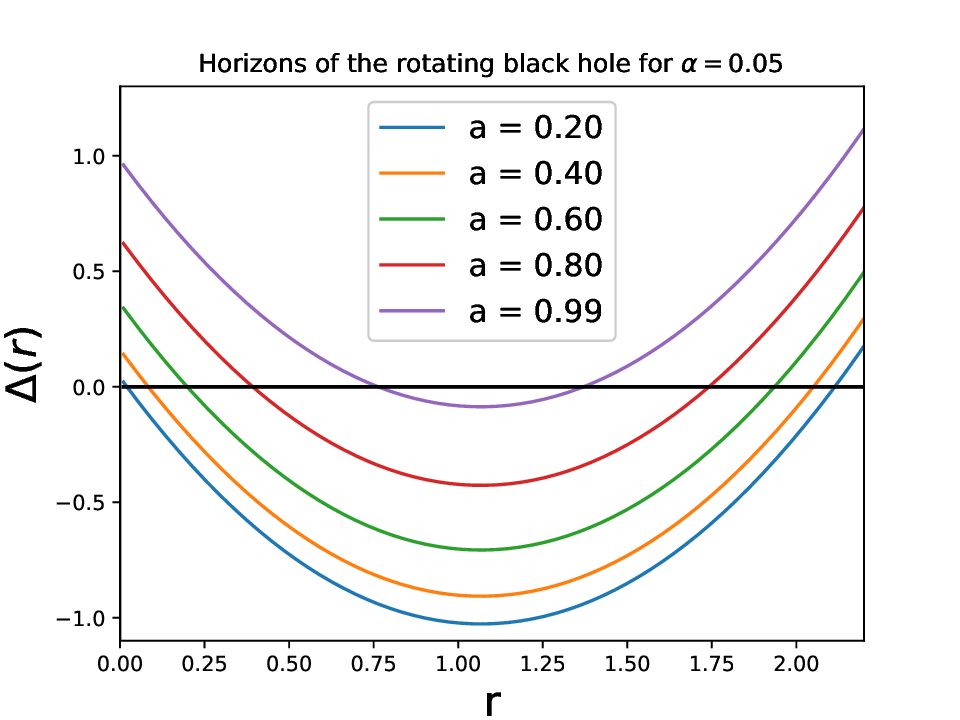}
    \includegraphics[width=0.4\linewidth]{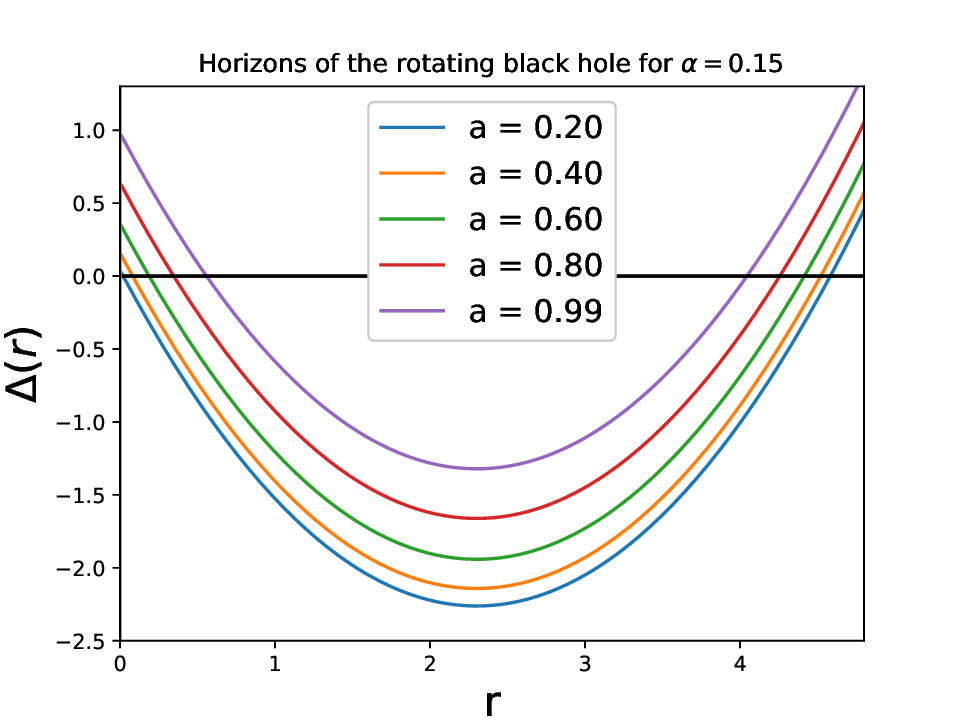}
    \caption{Variation of black hole horizons for different values of $\alpha$ and $a$}
    \label{fig1}
\end{figure}

\begin{figure}
    \centering
    \includegraphics[width=0.4\linewidth]{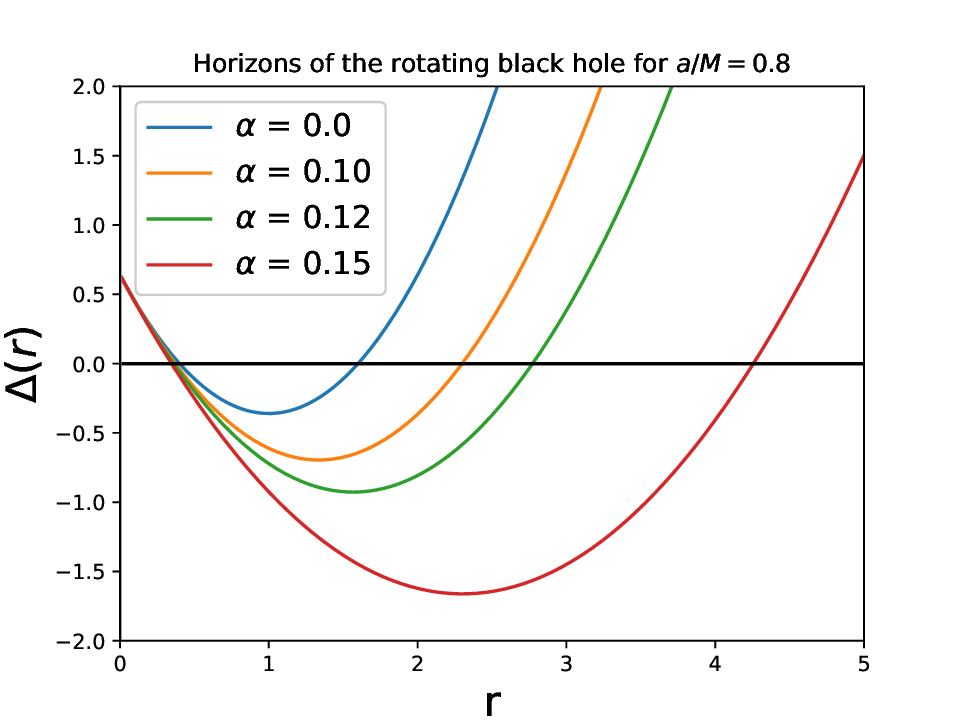}
    \includegraphics[width=0.4\linewidth]{dlt5.eps}
    \caption{Variation of black hole horizons for different values of $\alpha$ and $a$.}
    \label{fig2}
\end{figure}

\begin{figure}
    \centering
    \includegraphics[width=0.45\linewidth]{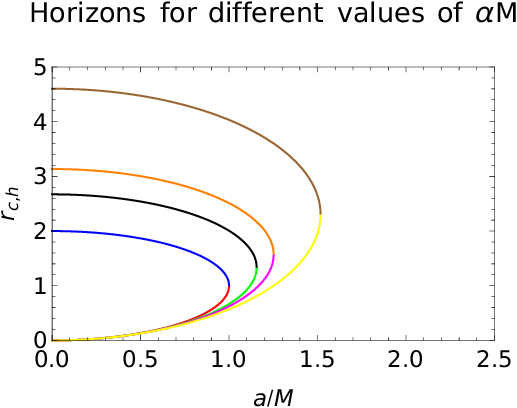}
    \includegraphics[width=0.5\linewidth]{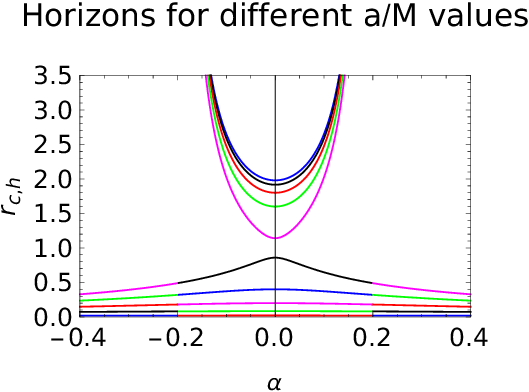}\\
    \caption{Variation of black hole horizons for different values of $\alpha$ and $a$.}
    \label{horizon_radii}
\end{figure}

{\label{metric}}
The spherically symmetric solution of the black hole in the Boyer-Lindquist coordinates is given by \cite{Dadhich:1997mh, Singha:2025huj}
\begin{eqnarray}\label{01}
ds^2=-f(r)dt^2+f^{-1}(r)dr^2+ r^2(d\theta^{2}+ \sin^{2}\theta d\phi^{2}),
\end{eqnarray}
with
\begin{equation}
    f(r) = 1- \frac{2M}{r} - 8\pi\alpha^2,
\end{equation}
where $M$ is the black hole mass, $r$ is the radial coordinate and $\alpha$ is the global monopole charge parameter. The event horizon of the black hole can be obtained by simply solving $g_{tt} = 0$ which gives
\begin{equation}
r = -\frac{2 M}{8 \pi  \alpha ^2-1}.
\end{equation}
The black hole corresponds to a single horizon, and in the absence of global monopole charge parameter, it reduces to the event horizon for the Schwarzschild black hole $2M$.
Now, by applying the Newman–Janis algorithm, one can find the rotating black hole metric and in the Boyer-Lindquist coordinates, the metric is given by \cite{Singha:2025huj}
\begin{equation}
ds^2 = -\left(1 - \frac{2Mr}{\Sigma} \right) dt^2 
- \frac{4Mar \sin^2\theta}{\Sigma} \, dt \, d\phi 
+ \frac{\Sigma}{\Delta} \, dr^2 
+ \Sigma \, d\theta^2 
+ \left(r^2 + a^2 + \frac{2Ma^2 r \sin^2\theta}{\Sigma} \right)\sin^2\theta \, d\phi^2,
\end{equation}
where 
\begin{equation}
\Sigma = r^2 + a^2 \cos^2\theta, \quad \text{and} \quad
\Delta = r^2 - 2Mr + a^2 - 8 \pi \alpha^2 r^2.
\end{equation}
The black hole corresponds to two horizons as
\begin{equation}
   r_{\pm} =  \frac{M \pm \sqrt{8 \pi  \alpha ^2 a^2-a^2+M^2}}{1-8 \pi  \alpha ^2},
\end{equation}
where $r_{\pm}$ corresponds to the inner and outer horizons of the black hole. In Fig.~({\ref{fig1}}), ({\ref{fig2}}) and Fig.~({\ref{horizon_radii}}), we plot the black hole horizons for different values of spin parameter $a$ and global monopole charge parameter $\alpha$. This metric has the similar structure as of the Kerr spacetimes. It is a axisymmetric metric having two Killing vectors $\xi^{\mu}_{(t)}$ and $\chi^{\mu}_{(\phi)}$. The vector $\xi^{\mu}_{(t)}$ corresponds to the time translation symmetry while the vector $\chi^{\mu}_{(\phi)}$ corresponding to the axial symmetry. Since the metric is not symmetric only under $t\to -t$, reflecting the fact that the Kerr black hole in the presence of global monopole charge is not static but stationary. The linear combination, $$\zeta^{\mu}=\xi^{\mu}_{(t)}+\Omega_h\chi^{\mu}_{(\phi)}$$, of the Killing vector fields $\xi^{\mu}_{(t)}$ and $\chi^{\mu}_{(\phi)}$ is also a Killing field, where $\Omega_h$ is the angular velocity at the event horizon and of course, $\zeta^{\mu}$ satisfies the Killing equation. 
\section{Thermodynamics of the Black hole}
\label{thermodynamics}
\begin{figure}
    \centering
    \includegraphics[width=0.48\linewidth]{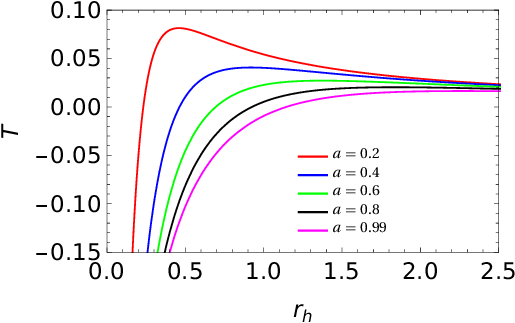}
    \includegraphics[width=0.48\linewidth]{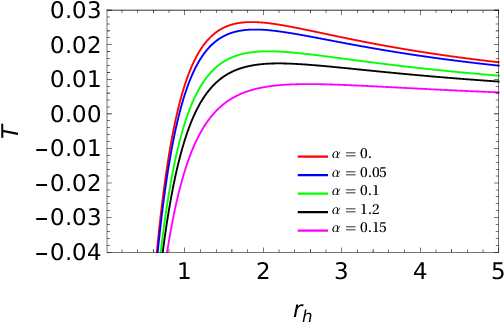}\\
    \caption{Variation of temperature with respect to the horizon radius for different values of $\alpha$ and $a$.}
    \label{Horizon_temp}
\end{figure}

In this section, we devote ourselves to a discussion of black hole thermodynamics. The thermal properties of the black hole are one of the exotic features of black hole mechanics. As in the usual thermal systems, we do calculate the temperature, entropy and other thermodynamic quantities, the black hole mechanics also resemble those properties in the geometric sense. We start with the definition of the surface gravity which expressed as $\kappa_h=-\frac{1}{2}\nabla_\mu\zeta_\nu \nabla^\mu\zeta^\nu|_{r=r_h}$. The temperature of the black hole event horizon is related to the surface gravity, via., $T=\frac{\kappa_h}{2\pi}$. To calculate the temperature, we first calculate the mass of the black hole by requiring that $\Delta(r_h)=0$. Therefore mass of the black hole reads as 

\begin{eqnarray}
    \label{mass}
M=\frac{a^2+r^2}{2 r_h}-{4 \pi  \alpha^2 r_h}.
\end{eqnarray}
Due to the presence of the global monopole charge, the mass of the black hole is modified. Since $\alpha>0$, the global monopole term contributes negatively to the mass of the black hole. Hence effective mass should be less than the usual Kerr black hole mass in the presence of the global monopole charge. This might be a reflection of the fact that global monopole charge effectively reduces the effect of gravitation as experienced by the objects near the event horizon of the Kerr black hole. \\
Next, we evaluate the temperature in terms of the metric coefficients, we have $T=\frac{1}{4\pi}\frac{\Delta^\prime(r_h)}{r_h^2+a^2}$ which upon putting the values of $\Delta(r)$ at the event horizon $r=r_h$, yields that
\begin{eqnarray}
\label{temp}
T=T_{\text{Kerr}}-\frac{2\alpha ^2 r_h}{a^2+r_h^2},
\end{eqnarray}
where $T_{\text{Kerr}}=\frac{1}{4 \pi }\frac{r_h^2-a^2}{a^2 r_h+r_h^3}$ is the Kerr black hole temperature. The global monopole modified Kerr black hole has a lower temperature than the original Kerr black hole, and hence we can say that modified black hole is cooler than the Kerr one. We plot in Fig.~\ref{Horizon_temp}, the temperature with respect to the horizon temperature, and we can see that for a fixed value of the global monopole charge, we have the maxima of the temperature profile corresponding to different values of the rotation parameter $a$. We can see from the figure that with increasing values of the rotation parameter, the maxima of the profile correspond to the value with lower temperatures. Similar is the situation for the right of Fig.~\ref{Horizon_temp}, where we fix the value of the rotation parameter and vary the global monopole charge corresponding to different temperature profiles. 
\begin{figure}
    \centering
    \includegraphics[width=0.48\linewidth]{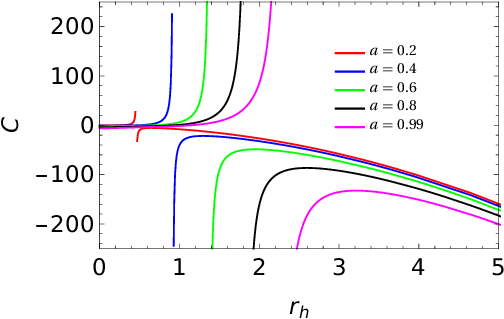}
    \includegraphics[width=0.48\linewidth]{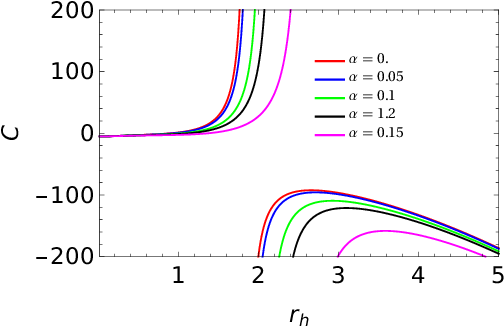}\\
    \caption{Variation of the heat capacity with respect to the horizon radius for different values of $\alpha$ and $a$.}
    \label{heat_cap}
\end{figure}
Next, we calculate the heat capacity to know the nature of the local stability of the black hole. The expression for the heat capacity is written as
\begin{eqnarray}
    \label{heat_capacity}
C=T\left(\frac{\partial S}{\partial T}\right)_{a,\alpha}=\frac{2 \pi  \left(a^2+r^2\right) \left(a^2+\left(8 \pi  \alpha ^2-1\right) r_h^2\right)}{\left(8 \pi  \alpha ^2-1\right) r_h^2-a^2},
\end{eqnarray}
We can see from the expression of the heat capacity that it diverges at a point where the horizon radius has the value $r_h=\frac{a}{\sqrt{8\pi\alpha^2-1}}$, and consequently, the temperature becomes maximum there. We plot in Fig.~\ref{heat_cap} the heat capacity with respect to different values of the rotation parameter $a$ (left) and with respect to different values of the global monopole charge $\alpha$ (right). We know that the negative (positive) heat capacity corresponds to the thermodynamically local (in)stability of the black holes. At the phase-transition point, the heat capacity diverges.\\
The global stability of the black hole is determined by the Gibbs free energy. We can express it as $G=M-TS$, which in terms of the black hole horizon radius and the other parameters as follows
\begin{eqnarray}
\label{Gibbs}
G=\frac{3 a^2}{4 r}-2 \pi  \alpha ^2 r+\frac{r}{4}.
\end{eqnarray}
\begin{figure}
    \centering
    \includegraphics[width=0.48\linewidth]{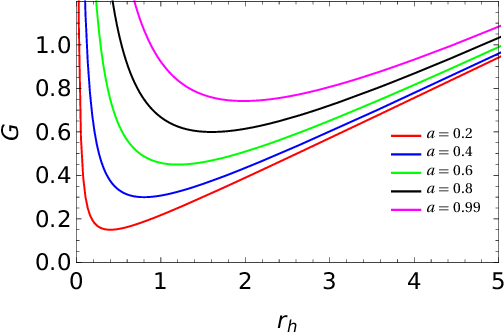}
    \includegraphics[width=0.48\linewidth]{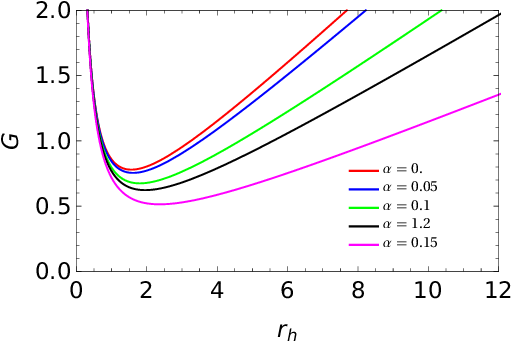}\\
    \caption{Variation of the Gibbs free energy with respect to the horizon radius for different values of $\alpha$ and $a$.}
    \label{Gibbs_free}
\end{figure}
One can see from Fig.~\ref{Gibbs_free}, the behavior of the Gibbs free energy with respect to the horizon radius. It is always positive, and hence reflecting the fact that the black hole is globally unstable. 
\section{Geodesic equation and Effective potential} 
\label{geodesics}
\begin{figure}
    \centering
    \includegraphics[width=0.45\linewidth]{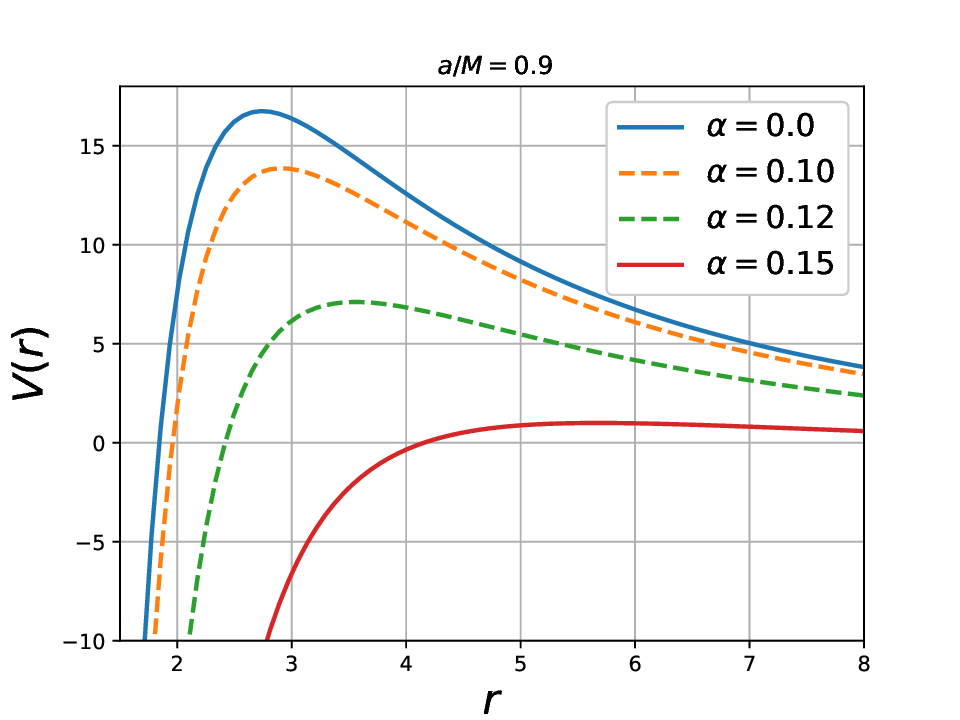}
     \includegraphics[width=0.45\linewidth]{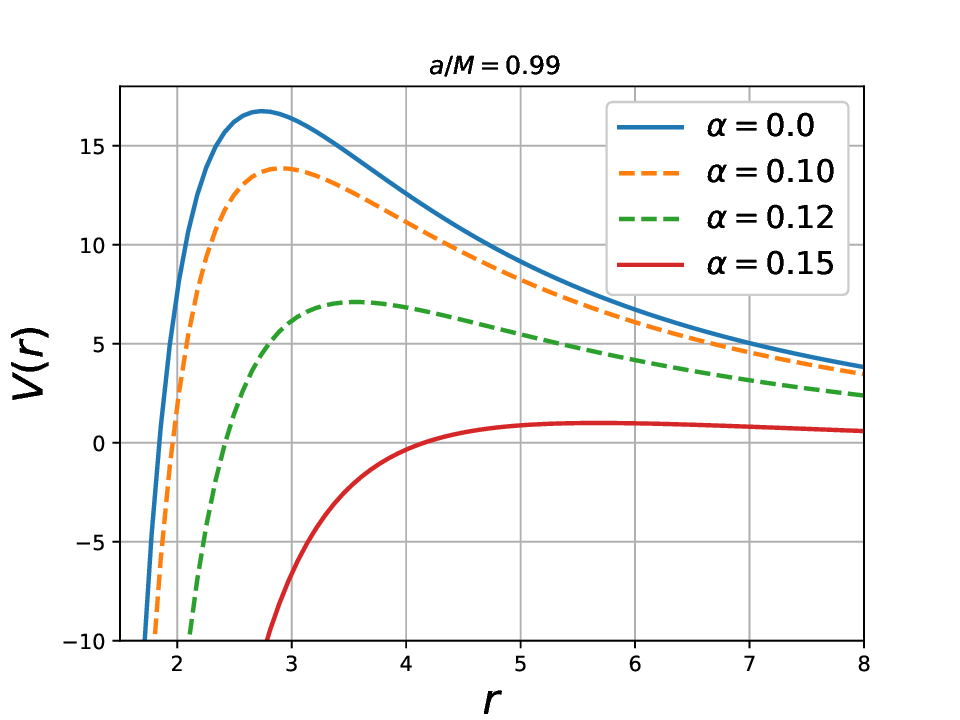}\\
     \includegraphics[width=0.45\linewidth]{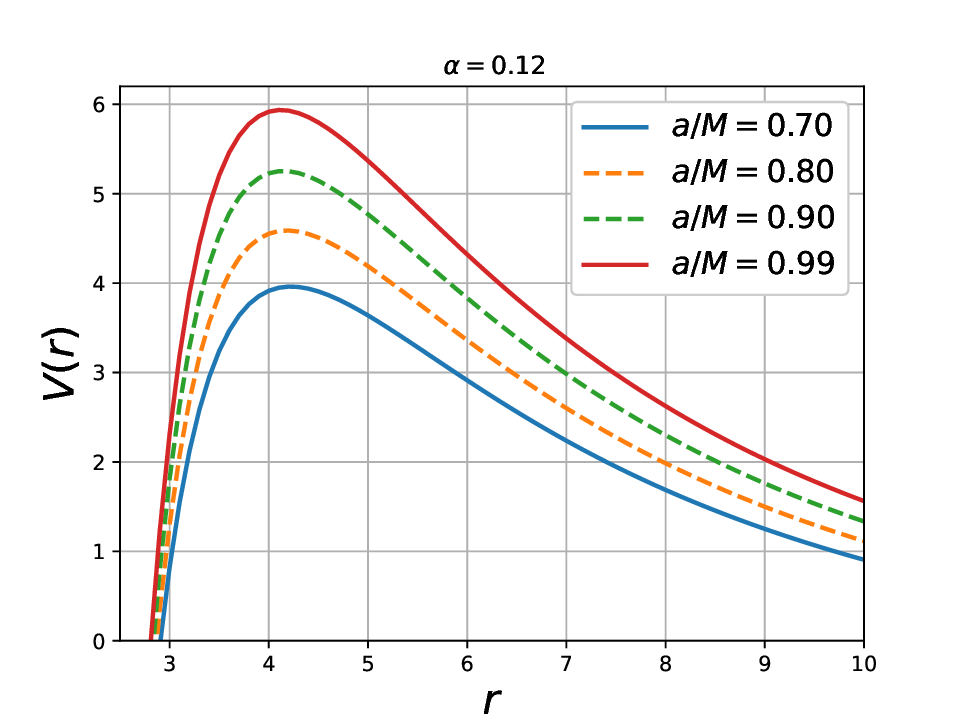}
     \includegraphics[width=0.45\linewidth]{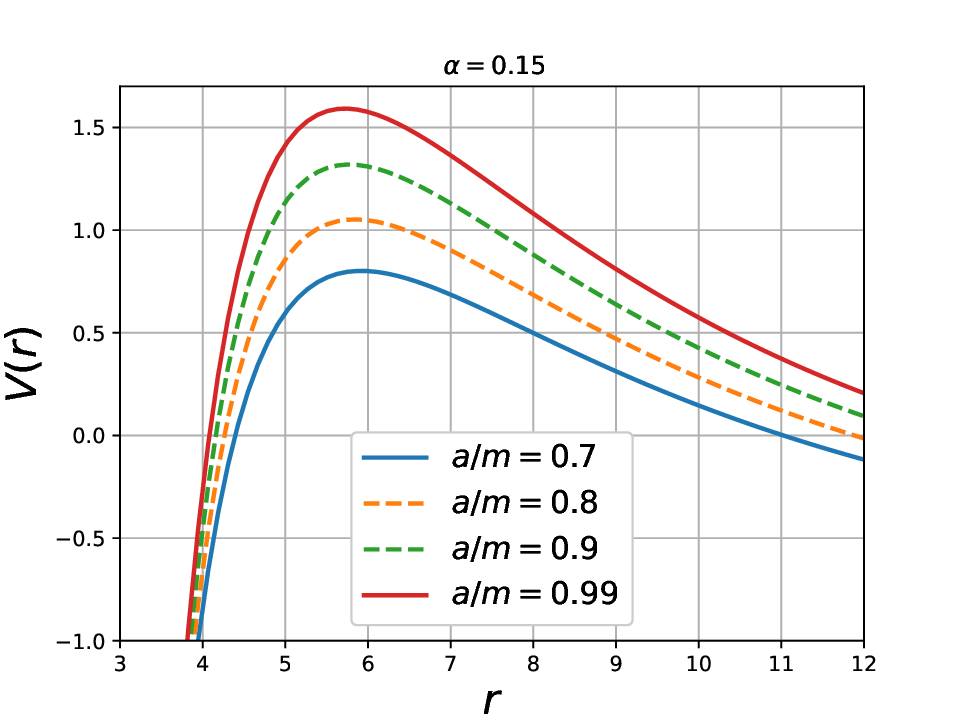}
    \caption{Effective potential of the black hole for different values of black hole spin $a$ and global monopole charge parameter $\alpha$.}
    \label{fig7}
\end{figure}
In this section, we derive the geodesic equations of motion for a test particle in the orbits of the black hole. We consider the Hamilton-Jacobi and Carter separable method to derive the entire set of equations of motion \cite{Chandrasekhar:1985kt, Carter:1971zc}. The Hamilton-Jacobi equation for motion is given by
\begin{equation}
    \mathcal{H} = -\frac{1}{2}g^{\mu\nu}\frac{\partial S}{\partial x^{\mu}} \frac{\partial S}{\partial x^{\nu}},
\end{equation}
where $S$ is the black hole action. The above equation is solvable by using a separable solution, and here we consider a solution such as \cite{Carter:1971zc}
\begin{eqnarray}
\label{eqn9}
S=\frac12 {m_{0}}^2 \tau -{ E} t +{ L} \phi + S_r(r)+S_\theta(\theta) ,
\end{eqnarray}
where $S_r(r)$ and $S_{\theta}(\theta)$ are functions of $r$ and $\theta$ only, and $m_{0}$ is the rest mass of the test particle, for the case of photon, the rest mass is zero. The differential equations obtained from the Hamilton-Jacobi equations are completely integrable, and the complete geodesic equations for a test particle can be obtained by solving the Hamilton-Jacobi equation with the help of a separable solution. The derived geodesic equations are given as \cite{Carter:1971zc}
\begin{eqnarray}
\Sigma \frac{dt}{d\tau}&=&\frac{r^2+a^2}{r^2 - 2Mr + a^2 - 8 \pi \alpha^2 r^2}\left[{ E}(r^2+a^2)-a{ L}\right]  -a(a{ E}\sin^2\theta-{ {L}})\ ,\label{t}\\
\Sigma \frac{dr}{d\tau}&=&\pm\sqrt{\mathcal{R}(r)}\ ,\label{r}\\
\Sigma \frac{d\theta}{d\tau}&=& \pm \sqrt{\Theta(\theta)}\ ,\label{th}\\
\Sigma \frac{d\phi}{d\tau}&=&\frac{a}{r^2 - 2Mr + a^2 - 8 \pi \alpha^2 r^2}\left[{ E}(r^2+a^2)-a{\cal L}\right]-\left(a{ E}-\frac{{ L}}{\sin^2\theta}\right)\ ,\label{phi}
\end{eqnarray}
where
\begin{eqnarray}
&&\mathcal{R}(r)=\left[(r^2+a^2){ E}-a{ L}\right]^2-\Delta(r)\left[{m_0}^2r^2+(a{ E}-{ L})^2+{ K}\right],\\
&&\Theta(\theta)={ K}-\left[\frac{{ L}^2}{\sin^2\theta}-a^2 { E}^2\right]\cos^2\theta\ .
\end{eqnarray}
and $K$ is the Carter separable constant \cite{Carter:1971zc}. The above geodesic equations describe the dynamics of a test particle around a rotating black hole with a global monopole charge. In the absence of the global monopole charge parameter, these equations reduce to the Kerr black hole. For the stability analysis of the particle in the gravitational field of the black hole, we require the effective potential of the black hole and which can be obtained by the radial equation of motion, and the derived effective potential is given by \cite{Frolov:2014dta}
\begin{equation}
    V_{eff}=\frac{1}{{{\Sigma}^2}}[ ((r^2+a^2) -a{\xi})^2-\Delta(r) ( (a -{\xi})^2+{\eta}) ].\label{vef}
\end{equation}
where $ \eta={K}/{E}^2$ and $\xi = {L}/{E}$ are the impact parameters. We plot the effective potential $V(r)$ with respect to the radial coordinate $r$ in Fig.~\ref{fig7}. For a fixed value of spin parameter $a$, the peak of the effective potential shifts to the right for increasing values of the global charge parameter $\alpha$, which emphasizes that the critical radius $r_c$ for forming the unstable orbits around the black hole increases with $\alpha$.   To obtain the unstable orbits of test particles around the black hole, we are maximizing the effective potential that must follow the condition \cite{Frolov:2014dta}
\begin{equation}
V_{eff}=\frac{\partial V_{eff}}{\partial r}=0 \;\; \;\; \mbox{or}\;\;  \; \mathcal{R}=\frac{\partial \mathcal{R}}{\partial r}=0,\label{20} 
\end{equation}
and by simultaneously solving the above equations, we obtain the separate equations for the impact parameters given as \cite{Chandrasekhar:1985kt}
 \begin{eqnarray} \label{eta}
 \eta&=& \frac{r^3 \left(4 a^2 M-r \left(-3 M-8 \pi  \alpha ^2 r+r\right)^2\right)}{a^2 \left(M+\left(8 \pi  \alpha ^2-1\right) r\right)^2},
 \end{eqnarray}
\begin{eqnarray} \label{xi}
\xi&=& \frac{a^2 M+8 \pi  a^2 \alpha ^2 r+a^2 r-3 M r^2-8 \pi  \alpha ^2 r^3+r^3}{a \left(M+8 \pi  \alpha ^2 r-r\right)}.
\end{eqnarray}
These equations of $\eta$ and $\xi$ describe photons motion around the rotating black hole with global monopole charge. The Eq.~(\ref{eta}) and Eq.~(\ref{xi}) reduce for the Kerr black holes in the absence of global monopole charge $\alpha = 0$.
\section{Black Hole Shadow}
{\label{shadow}}
\begin{figure}
    \centering
    \includegraphics[width=0.4\linewidth]{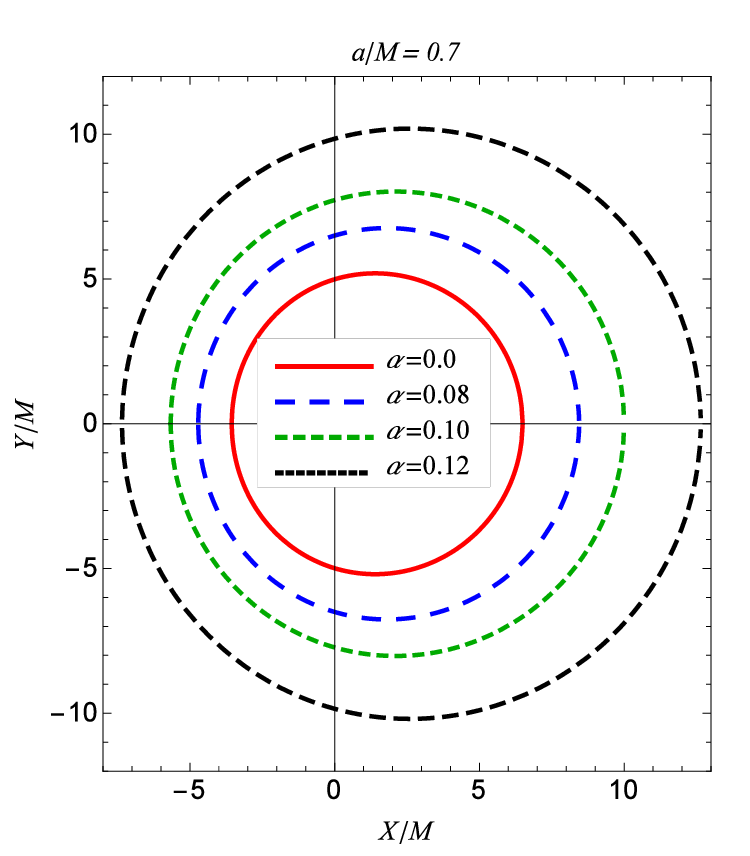}
    \includegraphics[width=0.42\linewidth]{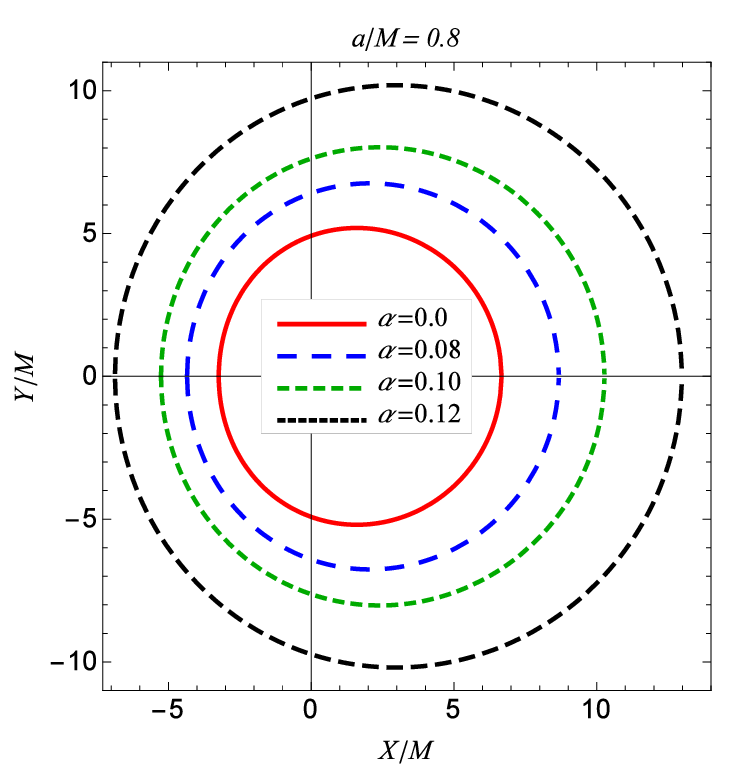}\\
    \includegraphics[width=0.4\linewidth]{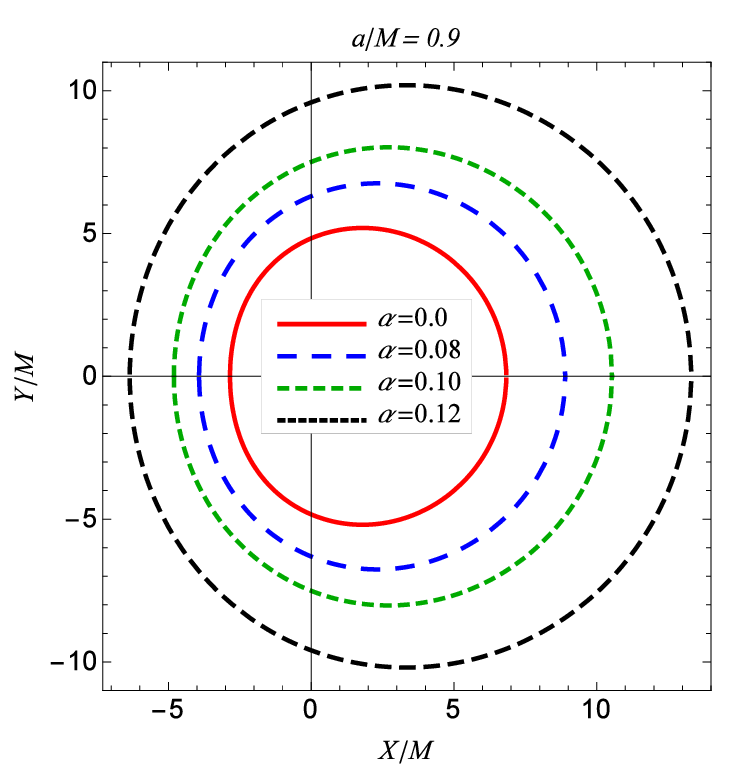}
    \includegraphics[width=0.4\linewidth]{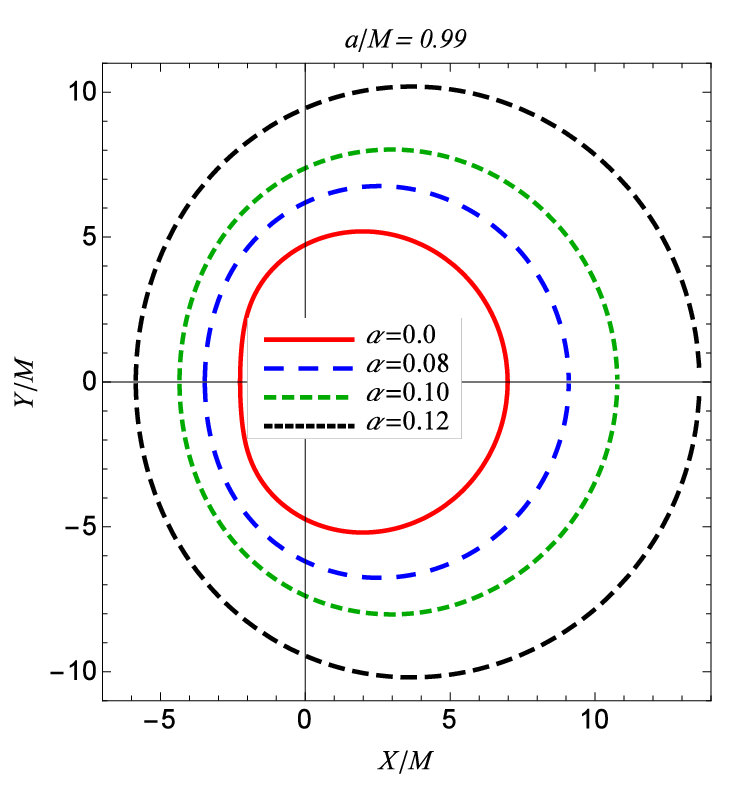}
    \caption{Plot showing the black hole shadow with the variation of spin parameter $a$ and global monopole charge parameter $\alpha$ and for $\theta_{0} = \pi/2$.}
    \label{fig8}
\end{figure}
\begin{figure}
    \centering
    \includegraphics[width=0.45\linewidth]{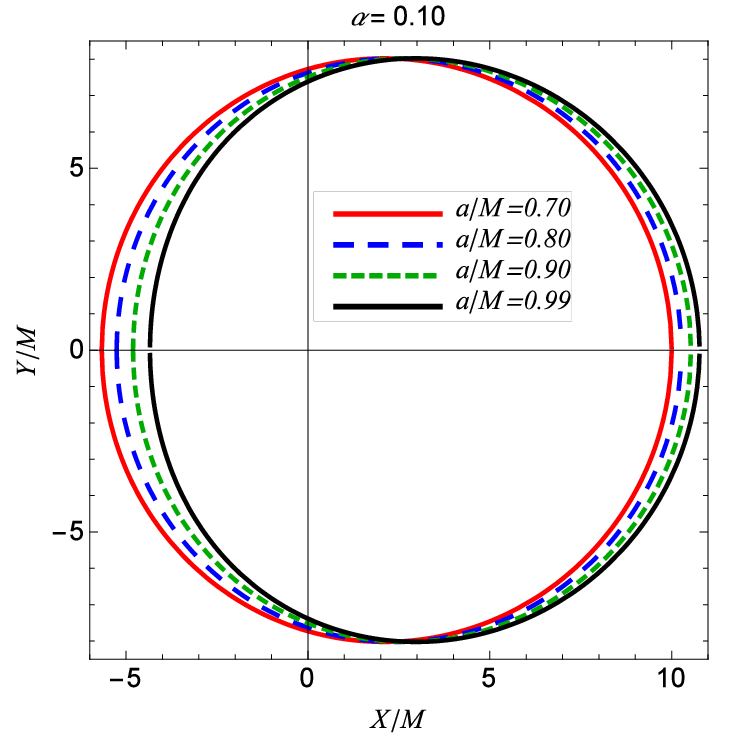}
    \includegraphics[width=0.42\linewidth]{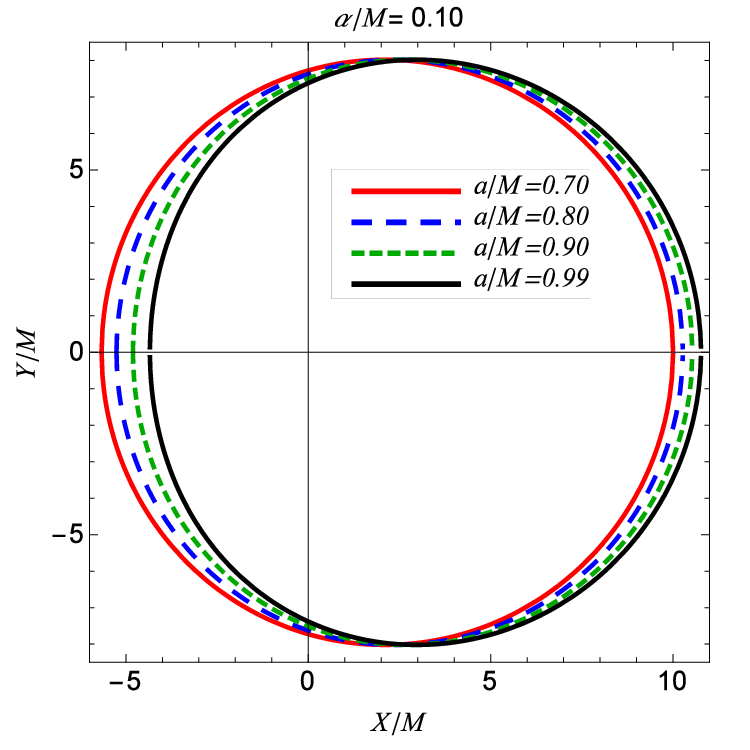}\\
    \includegraphics[width=0.43\linewidth]{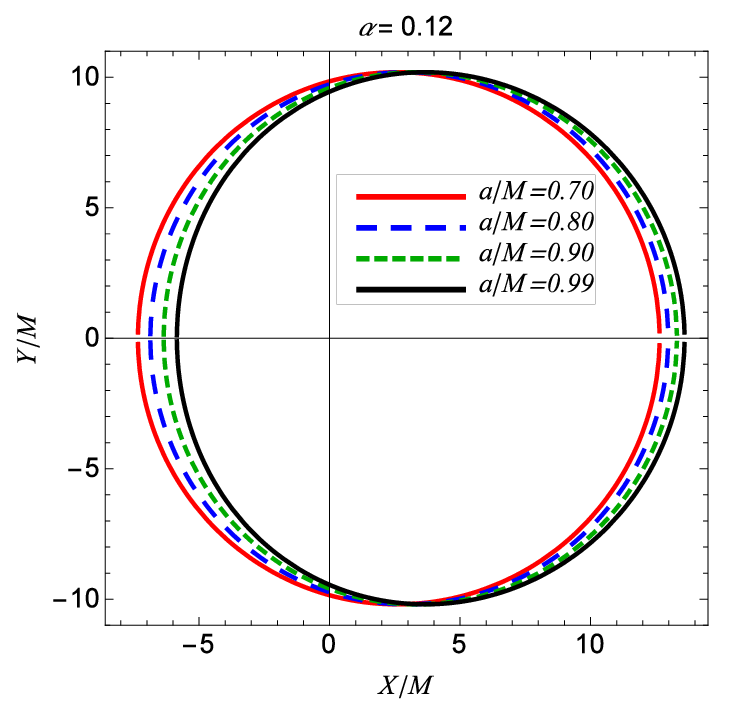}
    \includegraphics[width=0.43\linewidth]{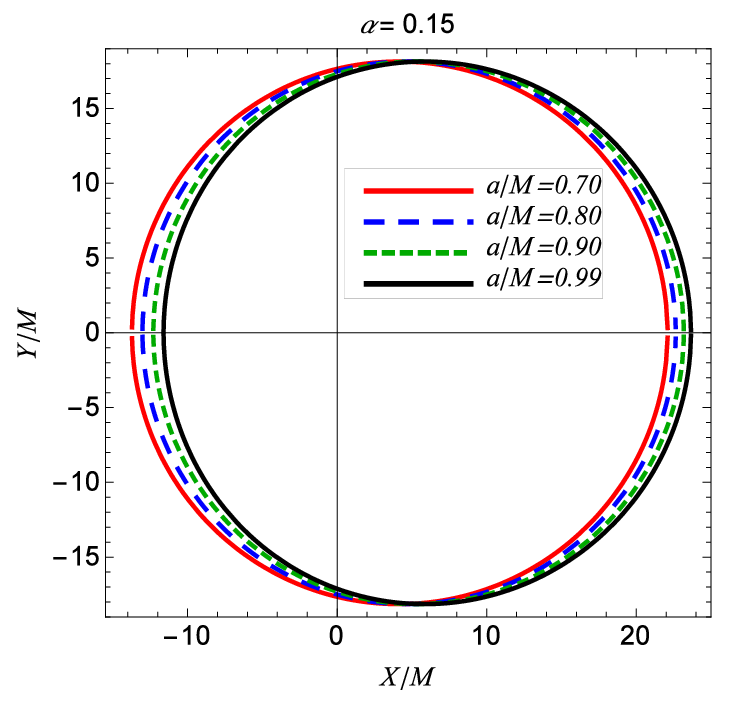}
    \caption{Plot showing the black hole shadow with the variation of spin parameter $a$ and global monopole charge parameter $\alpha$ and for $\theta_{0} = \pi/2$.}
    \label{fig9}
\end{figure}
In this section, we study the black hole shadow with the effect of global monopole charge. For better visualization of the black hole shadow, it is required to define celestial coordinates. The celestial coordinates with the relation of impact parameters $(\eta, \xi)$ are given by \cite{bardeen1973}
\begin{align}
X=&\lim_{r_0\rightarrow\infty}\left(-r_0^2 \sin{\theta_0}\frac{d\phi}{d{r}}\right),\nonumber\\ Y=&\lim_{r_0\rightarrow\infty}\left(r_0^2\frac{d\theta}{dr}\right),\label{Celestial1}
\end{align} 
for an infinitely far observer the celestial coordinate Eq.~(\ref{Celestial1}) leads to
\begin{align}
X=&-\xi{\csc\theta_0},\nonumber\\
Y=&\pm\sqrt{\eta+a^2\cos^2\theta_0-\xi^2\cot^2\theta_0}\ .\label{Celestial2}    
\end{align}
that must satisfy
\begin{equation}\label{xy}
X^2+Y^2=\eta+\xi^2+a^2\cos^2\theta_0.
\end{equation}
At the equator of the black hole, the Eq.~(\ref{xy}),  takes the form
\begin{equation}
    X^2 + Y^2 = \frac{a^2 \left(M+8 \pi  \alpha ^2 r+r\right)^2+2 r^2 \left(r^2-\left(M+8 \pi  \alpha ^2 r\right) \left(3 M+8 \pi  \alpha ^2 r\right)\right)}{\left(M+\left(8 \pi  \alpha ^2-1\right) r\right)^2}.
\end{equation}
The contour of the above Eq.~(\ref{xy}) traces the rotating black hole shadow with the global monopole charge. In Fig.~(\ref{fig8}) and Fig.~(\ref{fig9}), we plot the black hole shadow for different values of the spin parameter $a$ and the global monopole charge parameter $\alpha$. The plots in the  Fig.~(\ref{fig8}) show the black hole shadow for fixed spin parameter $a = 0.7$ and different values of the global charge parameter $\alpha$. In the Fig.~(\ref{fig8}), the first red color contour corresponds to the shadows of the Kerr black hole, and as we increase the global charge parameter, the effective size of the black hole shadow increases, and that is very clear from the fourth plot where we maximize the spin parameter $a = 0.99 M$. Here, for increasing values of the global charge parameter, the black hole shadow appears to be more circular (cf.~Fig.~\ref{fig8} and \ref{fig9}). 
\begin{table}[]
\begin{tabular}{|cccccccccc|}
\hline
\multicolumn{10}{|c|}{Radius and Distortion parameter}                                                                                                                                                                                                                                                                                   \\ \hline
\multicolumn{1}{|c|}{{S. No.}} & \multicolumn{1}{c|}{{$\alpha$}} & \multicolumn{2}{c|}{$a/M = 0.7$}                                 & \multicolumn{2}{c|}{$a/M = 0.8$}                                 & \multicolumn{2}{c|}{$a/M = 0.9$}                                 & \multicolumn{2}{c|}{$a/M = 0.99$}           \\ \cline{3-10} 
\multicolumn{1}{|c|}{}                        & \multicolumn{1}{c|}{}                          & \multicolumn{1}{c|}{$R_s$}   & \multicolumn{1}{c|}{$\delta_s$} & \multicolumn{1}{c|}{$R_s$}   & \multicolumn{1}{c|}{$\delta_s$} & \multicolumn{1}{c|}{$R_s$}   & \multicolumn{1}{c|}{$\delta_s$} & \multicolumn{1}{c|}{$R_s$}   & $\delta_s$ \\ \hline
\multicolumn{1}{|c|}{1}                       & \multicolumn{1}{c|}{0}                         & \multicolumn{1}{c|}{5.205}   & \multicolumn{1}{c|}{0.070629}   & \multicolumn{1}{c|}{5.205}   & \multicolumn{1}{c|}{0.091443}   & \multicolumn{1}{c|}{5.205}   & \multicolumn{1}{c|}{0.135322}   & \multicolumn{1}{c|}{5.205}   & 0.229016   \\ \hline
\multicolumn{1}{|c|}{2}                       & \multicolumn{1}{c|}{0.08}                      & \multicolumn{1}{c|}{6.75962} & \multicolumn{1}{c|}{0.059118}   & \multicolumn{1}{c|}{6.75962} & \multicolumn{1}{c|}{0.084712}   & \multicolumn{1}{c|}{6.75962} & \multicolumn{1}{c|}{0.109121}   & \multicolumn{1}{c|}{6.75962} & 0.142851   \\ \hline
\multicolumn{1}{|c|}{3}                       & \multicolumn{1}{c|}{0.12}                      & \multicolumn{1}{c|}{10.1944} & \multicolumn{1}{c|}{0.038391}   & \multicolumn{1}{c|}{10.1944} & \multicolumn{1}{c|}{0.052615}   & \multicolumn{1}{c|}{10.1944} & \multicolumn{1}{c|}{0.074097}   & \multicolumn{1}{c|}{10.1944} & 0.099896   \\ \hline
\multicolumn{1}{|c|}{4}                       & \multicolumn{1}{c|}{0.15}                      & \multicolumn{1}{c|}{18.1417} & \multicolumn{1}{c|}{0.03096}    & \multicolumn{1}{c|}{18.1417} & \multicolumn{1}{c|}{0.037024}   & \multicolumn{1}{c|}{18.1417} & \multicolumn{1}{c|}{0.043638}   & \multicolumn{1}{c|}{18.1417} & 0.0640332    \\ \hline
\end{tabular}
\caption{Table showing the variation of the observales, black hole radius $R_s$ and distortion parameter $\delta_s$ with different values of spin parameter $a$ and global magnetic charge parameter $\alpha$. }
\label{table1}
\end{table}
\begin{table}[]
\begin{tabular}{|cccccccccc|}
\hline
\multicolumn{10}{|c|}{Area and  Oblateness}                                                                                                                                                                                                                                                                                           \\ \hline
\multicolumn{1}{|c|}{{S. No.}} & \multicolumn{1}{c|}{{$\alpha$}} & \multicolumn{2}{c|}{$a/M = 0.7$}                             & \multicolumn{2}{c|}{$a/M = 0.8$}                             & \multicolumn{2}{c|}{$a/M = 0.9$}                             & \multicolumn{2}{c|}{$a/M = 0.99$}       \\ \cline{3-10} 
\multicolumn{1}{|c|}{}                        & \multicolumn{1}{c|}{}                          & \multicolumn{1}{c|}{$A/M^2$} & \multicolumn{1}{c|}{$D$}      & \multicolumn{1}{c|}{$A/M^2$} & \multicolumn{1}{c|}{$D$}      & \multicolumn{1}{c|}{$A/M^2$} & \multicolumn{1}{c|}{$D$}      & \multicolumn{1}{c|}{$A/M^2$} & $D$      \\ \hline
\multicolumn{1}{|c|}{1}                       & \multicolumn{1}{c|}{0}                         & \multicolumn{1}{c|}{82.1369} & \multicolumn{1}{c|}{0.969756} & \multicolumn{1}{c|}{81.0775} & \multicolumn{1}{c|}{0.956862} & \multicolumn{1}{c|}{79.5871} & \multicolumn{1}{c|}{0.932339} & \multicolumn{1}{c|}{77.1411} & 0.887002 \\ \hline
\multicolumn{1}{|c|}{2}                       & \multicolumn{1}{c|}{0.08}                      & \multicolumn{1}{c|}{139.843} & \multicolumn{1}{c|}{0.976312} & \multicolumn{1}{c|}{138.469} & \multicolumn{1}{c|}{0.966918} & \multicolumn{1}{c|}{136.679} & \multicolumn{1}{c|}{0.945439} & \multicolumn{1}{c|}{134.468} & 0.928574 \\ \hline
\multicolumn{1}{|c|}{3}                       & \multicolumn{1}{c|}{0.12}                      & \multicolumn{1}{c|}{197.552} & \multicolumn{1}{c|}{0.985004} & \multicolumn{1}{c|}{195.9}   & \multicolumn{1}{c|}{0.973693} & \multicolumn{1}{c|}{193.81}  & \multicolumn{1}{c|}{0.962952} & \multicolumn{1}{c|}{191.374} & 0.950052 \\ \hline
\multicolumn{1}{|c|}{4}                       & \multicolumn{1}{c|}{0.15}                      & \multicolumn{1}{c|}{320.296} & \multicolumn{1}{c|}{0.990262} & \multicolumn{1}{c|}{318.133} & \multicolumn{1}{c|}{0.981488} & \multicolumn{1}{c|}{315.468} & \multicolumn{1}{c|}{0.978181} & \multicolumn{1}{c|}{312.502} & 0.967983 \\ \hline
\end{tabular}
\caption{Table showing the variation of the observales area of the black hole shadow $A$ and oblatness $D$  with different values of spin parameter $a$ and global magnetic charge parameter $\alpha$. }
\label{table2}
\end{table}

\begin{figure}
    \centering
    \includegraphics[width=0.49\linewidth]{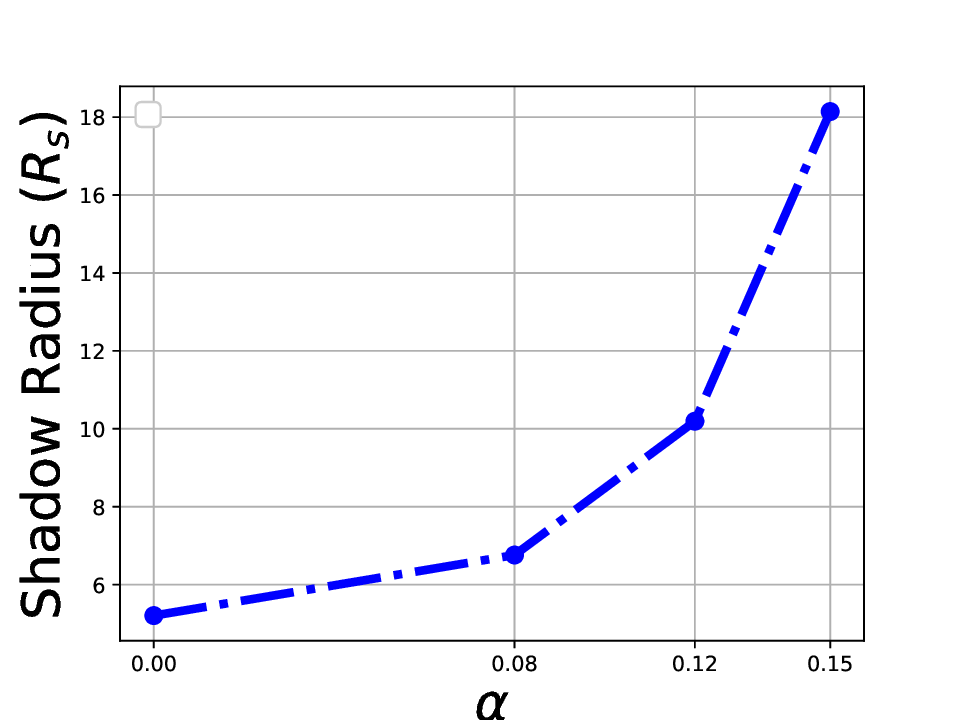}
    \includegraphics[width=0.49\linewidth]{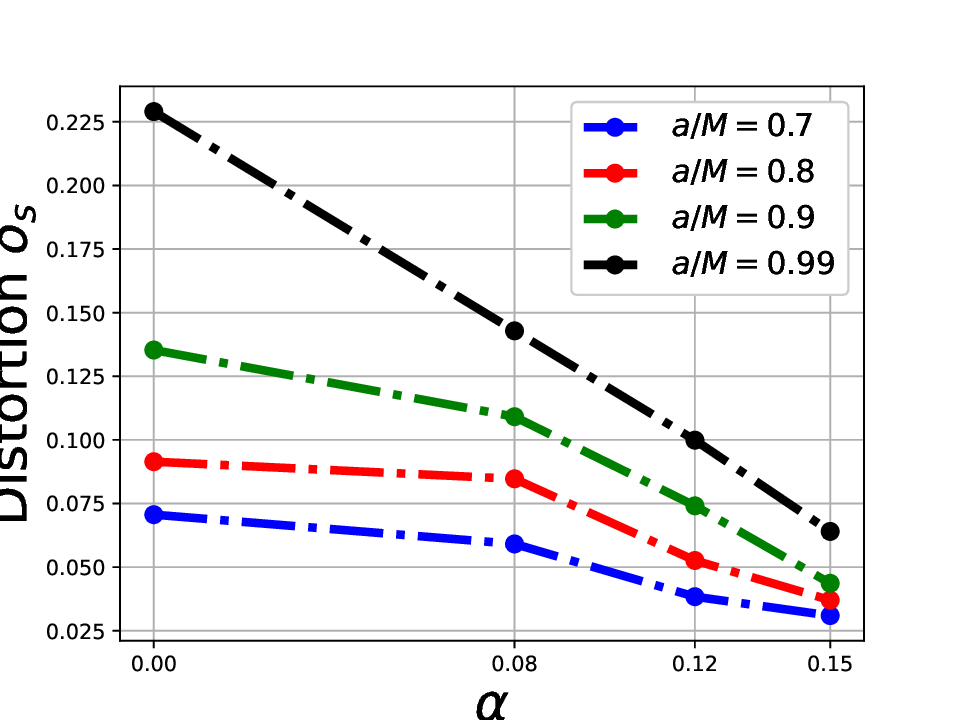}
\caption{Plot showing the variation of black hole shadow radius $R_s$ and Distortion $\delta_s$ with spin parameter $a$ and global monopole charge parameter $\alpha$.}
    \label{fig10}
\end{figure}

\begin{figure}
    \centering
   \includegraphics[width=0.49\linewidth]{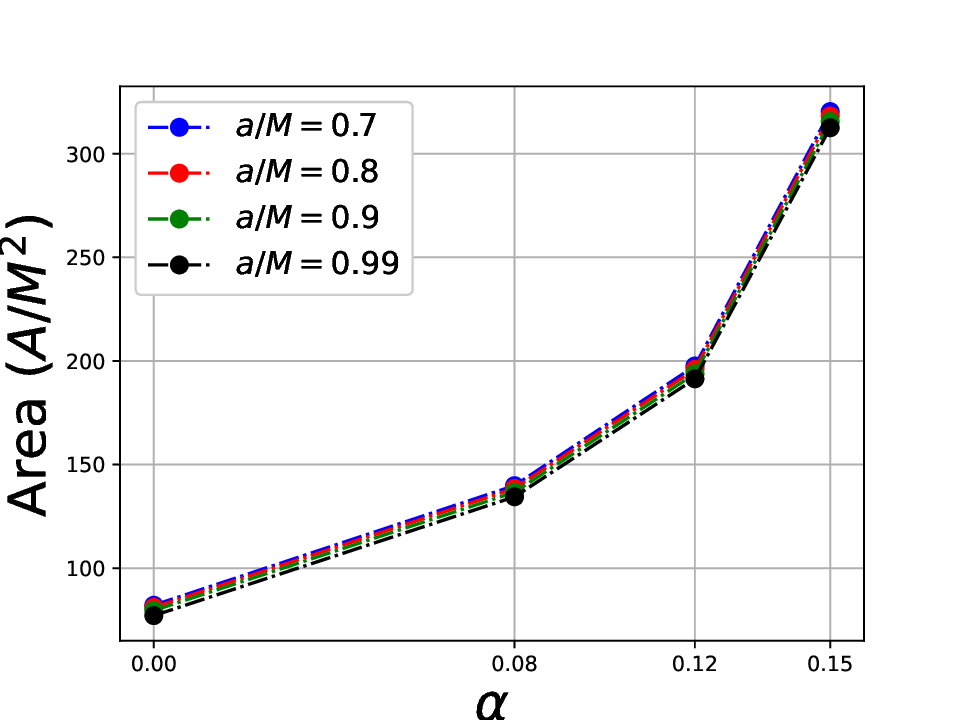}
    \includegraphics[width=0.49\linewidth]{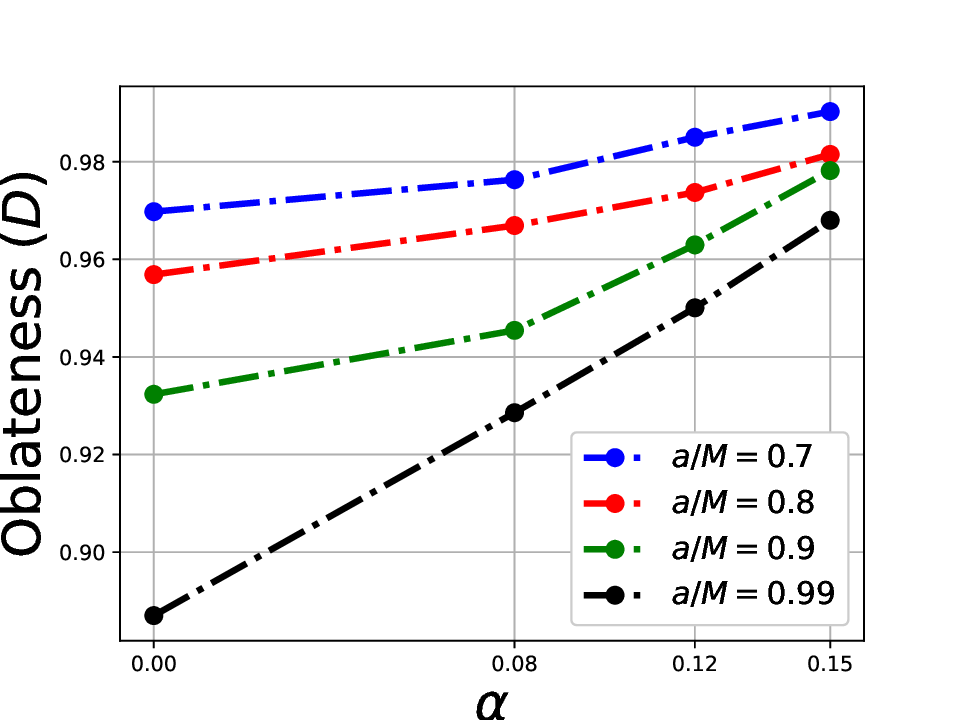}
\caption{Plot showing the variation of black hole shadow area $A$ and oblateness $D$ with spin parameter $a$ and global monopole charge parameter $\alpha$.}
    \label{fig11}
\end{figure}
The global charge parameter strongly affects the shape and size of the black hole shadow, and for a better understanding of the black hole shadow in the presence of the global charge parameter, we study the black hole observables that is its shadow radius $R_s$, the deformation parameter $\delta_s$, proposed by Hioki and Maeda \cite{Hioki:2009na} and other observables,  the area of the black hole shadow $A_s$ and the oblateness $D$, proposed by Kumar and Ghosh \cite{Kumar:2018ple}.
The geometry of the black hole shadow figures describes the shadow observables. The observable black hole shadow radius $R_s$ reads \cite{Hioki:2009na}
\begin{equation}
    R_s = \frac{(X_t - X_r)^2 + Y_t^2}{2 |X_r - X_t|},
\end{equation}
and the deformation parameter $\delta_s$ reads \cite{Hioki:2009na}
\begin{equation}
    \delta_s = \frac{\mathcal{D}}{R_s},
\end{equation}
where $\mathcal{D} = {x}^{'} - {x}$ is the dent appears in the black hole shadow. The next observable is the area under the black hole shadow curve, and it takes the following form \cite{Kumar:2018ple}
\begin{equation}
    A = 2 \int{Y}dX = 2 \int_{r_p^{-}}^{r_{p}^{+}} \left( Y \frac{dX}{dr_{p}} dr_{p}\right),
\end{equation}
and the other observable that we analytically measure takes the following form \cite{Kumar:2018ple}
\begin{equation}
    D = \frac{X_{r} - X_{t}}{Y_{t} - Y_{b}}.
\end{equation}
These observables measure the effective change in the shape and size of the black hole with the effect of the global magnetic charge parameter. We show the variation of the shadow radius $R_s$, and distortion $\delta_s$ in Fig.~({\ref{fig10}}) and Table~(\ref{table1}) and the variation of area $A$ and oblateness in Fig.~{\ref{fig11}} and Table~\ref{table2}. It's very clear from the Fig.~(\ref{fig10}) and (\ref{fig11}), and Table~(\ref{table1}) and (\ref{table2}), the shadow radius of the black hole increases, the distortion decreases, the area increases, and the oblateness $D$ increases with the increasing values of the global monopole charge parameter $\alpha$ with fixed $a$.
\section{Constraints with the EHT observations}
\label{Constraints}
In this section, we estimate the angular diameter of M87 and Sgr A$^{*}$ black hole shadow with the effect of spin parameter $a$, and global charge parameter $\alpha$, and check the consistency of these astrophysical black holes from the current EHT observational results. The angular diameter of the black hole shadow takes the following form \cite{Kumar:2020owy, Bambi:2019tjh}
\begin{equation}
    \theta_s = \frac{2}{r_{0}}\sqrt{\frac{A}{\pi}},
 \end{equation}
where $A$ is the area of the black hole shadow and $r_{0}$ is the distance of the black hole shadow to the observer at infinity. 
\subsection{M87 black hole}
The mass of the M87 black hole is $6.5 \times 10^{9}$ $M_{\odot}$ and its  $5.1839\times10^{20}$ $km$ far distant from Earth \citep{EventHorizonTelescope:2019dse, EventHorizonTelescope:2019pgp, EventHorizonTelescope:2019ggy,EventHorizonTelescope:2019jan, EventHorizonTelescope:2019ths, EventHorizonTelescope:2019uob}. Due to this very large distance from Earth,  the estimated size of the M87 black hole is in $\mu as$. In Table~{\ref{table3}}, we show the estimated values of the angular diameter of the M87 black hole for different values of the black hole spin $a$ and the global charge parameter $\alpha$. Our finding shows that for a fixed value of the spin parameter $a$, the angular diameter $\theta_s$ of the M87 black hole shadow increases consistently with increasing values of the global monopole charge parameter, as clearly shown by the columns of the Table.~{\ref{table3}} and Fig.~(\ref{fig12}). For fixed values of the global charge parameter $\alpha$, the angular diameter $\theta_s$ decreases with increasing values of the spin parameter $a$ as shown by the rows of Table~({\ref{table3}) and Fig.~(\ref{fig12}).
\begin{figure}
    \centering
    \includegraphics[width=0.8\linewidth]{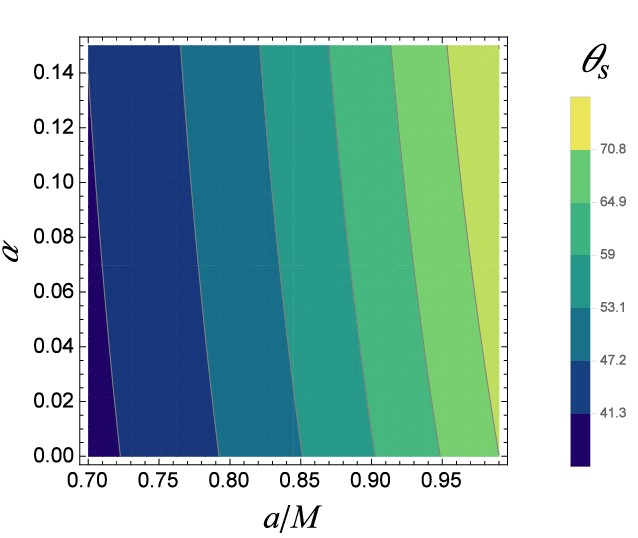}
    \caption{Variation of angular diameter of M87 black hole with black hole spin and global monopole charge parameter }
    \label{fig12}
\end{figure}
\begin{figure}
    \centering
    \includegraphics[width=0.8\linewidth]{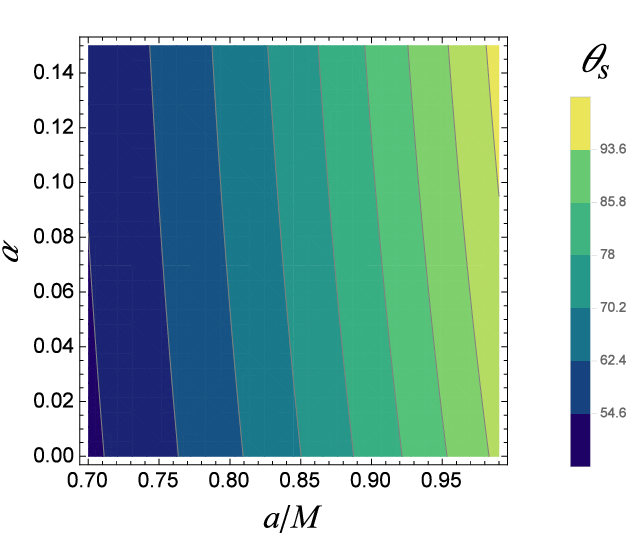}
    \caption{Variation of angular diameter of Sgr $A^{*}$ black hole with black hole spin and global monopole charge parameter }
    \label{fig13}
\end{figure}
\begin{table}[]
\begin{tabular}{|cccccc|}
\hline
\multicolumn{6}{|c|}{Angular   Diameter of M87 Black Hole}                                                                                                                                                            \\ \hline
\multicolumn{1}{|c|}{ {S.   No.}} & \multicolumn{1}{c|}{ {$\alpha$}} & \multicolumn{4}{c|}{$\theta_s(\mu s)$}                                                                                \\ \cline{3-6} 
\multicolumn{1}{|c|}{}                          & \multicolumn{1}{c|}{}                       & \multicolumn{1}{c|}{$a/M = 0.7$} & \multicolumn{1}{c|}{$a/M = 0.8$} & \multicolumn{1}{c|}{$a/M = 0.9$} & $a/M = 0.99$ \\ \hline
\multicolumn{1}{|c|}{1}                         & \multicolumn{1}{c|}{0}                      & \multicolumn{1}{c|}{39.22}       & \multicolumn{1}{c|}{38.97}       & \multicolumn{1}{c|}{38.61}       & 38.01        \\ \hline
\multicolumn{1}{|c|}{2}                         & \multicolumn{1}{c|}{0.08}                   & \multicolumn{1}{c|}{51.18}       & \multicolumn{1}{c|}{50.93}       & \multicolumn{1}{c|}{50.60}       & 50.19        \\ \hline
\multicolumn{1}{|c|}{3}                         & \multicolumn{1}{c|}{0.12}                   & \multicolumn{1}{c|}{60.83}       & \multicolumn{1}{c|}{60.58}       & \multicolumn{1}{c|}{60.25}       & 59.87        \\ \hline
\multicolumn{1}{|c|}{4}                         & \multicolumn{1}{c|}{0.15}                   & \multicolumn{1}{c|}{77.46}       & \multicolumn{1}{c|}{77.20}       & \multicolumn{1}{c|}{76.87}       & 76.51        \\ \hline
\end{tabular}
 \caption{Table showing the variation of angular diameter of M87 black hole with spin parameter $a$ and global monopole charge parameter $\alpha$. }
 \label{table3}
\end{table}
\begin{table}[]
\begin{tabular}{|cccccc|}
\hline
\multicolumn{6}{|c|}{Angular   Diameter of Sgr A$^{*}$ Black Hole}                                                                                                                 \\ \hline
\multicolumn{1}{|c|}{S.   No.} & \multicolumn{1}{c|}{$\alpha$} & \multicolumn{1}{c|}{$a = 0.7M$} & \multicolumn{1}{c|}{$a = 0.8M$} & \multicolumn{1}{c|}{$a = 0.9M$} & $a = 0.99M$ \\ \hline
\multicolumn{1}{|c|}{1}        & \multicolumn{1}{c|}{0}        & \multicolumn{1}{c|}{52.459}     & \multicolumn{1}{c|}{52.119}     & \multicolumn{1}{c|}{51.638}     & 50.838      \\ \hline
\multicolumn{1}{|c|}{2}        & \multicolumn{1}{c|}{0.01}     & \multicolumn{1}{c|}{52.66}      & \multicolumn{1}{c|}{52.321}     & \multicolumn{1}{c|}{51.840}    & 51.051     \\ \hline
\multicolumn{1}{|c|}{3}        & \multicolumn{1}{c|}{0.04}     & \multicolumn{1}{c|}{55.833}     & \multicolumn{1}{c|}{55.495}     & \multicolumn{1}{c|}{55.028}     & 54.345      \\ \hline
\multicolumn{1}{|c|}{4}        & \multicolumn{1}{c|}{0.08}     & \multicolumn{1}{c|}{68.45}      & \multicolumn{1}{c|}{68.112}     & \multicolumn{1}{c|}{67.671}     & 67.121      \\ \hline
\multicolumn{1}{|c|}{5}        & \multicolumn{1}{c|}{0.12}     & \multicolumn{1}{c|}{81.356}     & \multicolumn{1}{c|}{81.015}     & \multicolumn{1}{c|}{80.582}     & 80.074      \\ \hline
\multicolumn{1}{|c|}{6}        & \multicolumn{1}{c|}{0.15}     & \multicolumn{1}{c|}{103.593}    & \multicolumn{1}{c|}{103.242}    & \multicolumn{1}{c|}{102.809}    & 102.324     \\ \hline
\end{tabular}
 \caption{Table showing the variation of angular diameter of Sgr A$^{*}$ black hole with spin parameter $a$ and global monopole charge parameter $\alpha$. }
 \label{table4}
\end{table}

\subsection{Sgr A$^{*}$ black hole}
The Sgr A$^{*}$ black hole is situated at the heart of our galaxy, the Milky Way \citep{EventHorizonTelescope:2022exc,EventHorizonTelescope:2022urf,EventHorizonTelescope:2022apq,EventHorizonTelescope:2022wok,EventHorizonTelescope:2022wkp, EventHorizonTelescope:2022xqj}. The observed mass of this black hole is $4 \times 10^{6} M_{\odot}$ and is 8 $kps$ far from Earth. We show the estimated values of the angular diameter $\theta_s$ with the black hole spin parameter $a$ and the global charge parameter $\alpha$ in Table \ref{table4}.  The angular diameter of the Sgr A$^{*}$ black hole consistently increases with the increasing values of the global monopole charge parameter $\alpha$. It varies from $52.45 \mu as$ to $103.59 \mu as$ with the variation of $\alpha$  $\in$ (0, 0.15)   as clearly shown by all the rows of Table~(\ref{table4}). The values of black hole angular diameter significantly decrease with the spin parameter $a$ for fixed $\alpha$. It varies from $52.45 \mu as$ to $50.83 \mu as$ with the variation of $a$  $\in$ (0.7, 0.99) as clearly shown by all the columns of Table~(\ref{table4}). In Fig.~(\ref{fig13}), we plot the angular diameter of Sgr A$^*$ black hole shadow with spin parameter $a$ and global charge parameter $\alpha$. 
\section{conclusion}
\label{conclusion}
The study of black hole shadows is an effective way to find the non-existent properties of black holes. The current observations from EHT require examining general relativity as well as modified theories of gravity. In this article, we have shown that in the presence of global monopole charge $\alpha$, the black hole shadow effectively gets modified, and by considering the M87 and Sgr A$^{*}$ observations from EHT, we have obtained certain limits on the global monopole charge parameter $\alpha$. Our findings show the following results:
\begin{itemize}
    \item The event horizon of the black hole has been studied with the effect of the global monopole charge parameter. For a fixed value of spin parameter $a$, the inner horizon of the black hole shrinks and the outer horizon of the black hole expands for the increasing values of global charge parameter (cf.~Fig.~\ref {fig1}).
    \item All the thermodynamical properties of the black hole have been studied with the effect of the global charge parameter $\alpha$.
    \item The geodesic equations of motion have been originally derived with the effect of the global charge parameter $\alpha$, and the analysis of the effective potential has been performed for the different values of the global charge parameter $\alpha$ (cf.~Fig.~\ref {fig7}). 
    \item The study of the rotating black hole shadow has been done with the effect of global charge parameter $\alpha$, and for the increasing values of  $\alpha$, the black hole shadow appears larger and more circular (cf.~Fig.~\ref{fig8} and \ref{fig9} ).
    
    \item For better analysis of the black hole shadow, we obtained black hole observables such as shadow radius $R_s$, distortion parameter $\delta_s$, area $A$, and the oblateness $D$
    \item With the increasing values of the global charge parameter $\alpha$, the shadow radius $R_s$ monotonically increases, distorion parameter $\delta_s$ decreases, are $A$ increases and the oblateness $D$ decreases (cf.~Fig.~\ref {fig10} and Fig.~\ref {fig11}). Hence, the rotating black hole shadow appears larger and more circular in comparison to the Kerr black hole shadow.  
    \item We obtained the angular diameter of the black hole shadow with the current EHT observations for the astrophysical black holes M87 and Sgr A$^*$. 
    \item The angular diameter of the black hole shadow increases with the effect of global monopole charge parameter $\alpha$ for M87 and Sgr A$^*$ black holes (cf.~Fig.~\ref {fig12} and Fig.~\ref {fig13}). 
\end{itemize}
Till now, EHT has captured M87 and Sgr A{$^*$} black hole shadow. EHT is now improving its measurement techniques and upgrading to the Next Generation Event Horizon Telescope (ngEHT). We are expecting more astrophysical black hole shadow images from the ngEHT, and that will open more gateways to test GR as well as other theories of gravity in the very near future.
\\
\textbf{Credit authorship contribution statement}\\
Balendra Pratap Singh: Methodology, Software, Visualization, Investigation, Writing - original draft. \\
Md Sabir Ali: Conceptualization, Software,Investigation, Writing -original draft.\\
\textbf{Declaration of competing interest:}
The author declares that they have no known competing financial interests or personal relationships that could have appeared to influence the work reported in this paper.\\
\textbf{Data availability:}
No data have been used for the research described in the article.


\begin{thebibliography}{}

\bibitem{Schwarzschild:1916uq}
K.~Schwarzschild,
Sitzungsber. Preuss. Akad. Wiss. Berlin (Math. Phys. ) \textbf{1916}, 189-196 (1916)
[arXiv:physics/9905030 [physics]].

\bibitem{Reissner:1916cle}
H.~Reissner, 
\textit{Annalen der Physik}, vol.~355, no.~9, pp.~106--120 (1916).


\bibitem{Nordstrom:1918bv}
G.~Nordström, 
\textit{Koninklijke Nederlandsche Akademie van Wetenschappen Proceedings}, vol.~20, pp.~1238--1245 (1918).

\bibitem{Weyl:1917rtf}
H.~Weyl, 
\textit{Annalen der Physik}, vol.~359, no.~18, pp.~117--145 (1917).

\bibitem{Jeffery:1921rsl}
G.~B.~Jeffery, 
\textit{Proceedings of the Royal Society of London. Series A, Containing Papers of a Mathematical and Physical Character}, 
vol.~99, no.~697, pp.~123--134 (1921).

\bibitem{Kerr:1963ud}
R.~P.~Kerr, ``Gravitational field of a spinning mass as an example of algebraically special metrics,'' 
\textit{Physical Review Letters}, vol.~11, no.~5, pp.~237--238 (1963).

\bibitem{Newman:1965tw}
E.~T.~Newman and A.~I.~Janis, ``Note on the Kerr spinning-particle metric,'' 
\textit{Journal of Mathematical Physics}, vol.~6, no.~6, pp.~915--917 (1965).

\bibitem{Newman:1965my}
E.~T.~Newman, R.~Couch, K.~Chinnapared, A.~Exton, A.~Prakash, and R.~Torrence, 
``Metric of a rotating, charged mass,'' 
\textit{Journal of Mathematical Physics}, vol.~6, no.~6, pp.~918--919 (1965).

\bibitem{Barriola:1989hx}
M.~Barriola and A.~Vilenkin, 
``Gravitational field of a global monopole,'' 
\textit{Physical Review Letters}, vol.~63, no.~4, pp.~341--343 (1989).

\bibitem{Dadhich:1997mh}
N.~Dadhich, K.~Narayan and U.~A.~Yajnik,
Pramana \textbf{50}, 307-314 (1998)
[arXiv:gr-qc/9703034 [gr-qc]].

\bibitem{Bardeen:1973gs}
J.~M.~Bardeen, B.~Carter, and S.~W.~Hawking, 
\emph{Commun. Math. Phys.} \textbf{31}, 161--170 (1973).

\bibitem{Hawking:1975vcx}
S.~W.~Hawking, 
\emph{Commun. Math. Phys.} \textbf{43}, 199 (1975), 
[Erratum: \emph{Commun. Math. Phys.} \textbf{46}, 206 (1976)].

\bibitem{Bekenstein:1973ur}
J.~D.~Bekenstein, 
\emph{Phys. Rev. D} \textbf{7}, 2333--2346 (1973).

\bibitem{Singha:2025huj}
C.~Singha and N.~Dadhich,
[arXiv:2504.16949 [gr-qc]].

\bibitem{EventHorizonTelescope:2019dse}
K.~Akiyama \textit{et al.} [Event Horizon Telescope],
Astrophys. J. Lett. \textbf{875}, L1 (2019)
[arXiv:1906.11238 [astro-ph.GA]].

\bibitem{EventHorizonTelescope:2019pgp}
K.~Akiyama \textit{et al.} [Event Horizon Telescope],
Astrophys. J. Lett. \textbf{875}, no.1, L5 (2019)
[arXiv:1906.11242 [astro-ph.GA]].

\bibitem{EventHorizonTelescope:2019ggy}
K.~Akiyama \textit{et al.} [Event Horizon Telescope],
Astrophys. J. Lett. \textbf{875}, no.1, L6 (2019)
[arXiv:1906.11243 [astro-ph.GA]].

\bibitem{EventHorizonTelescope:2019jan}
K.~Akiyama \textit{et al.} [Event Horizon Telescope],
Astrophys. J. Lett. \textbf{875}, no.1, L3 (2019)
[arXiv:1906.11240 [astro-ph.GA]].

\bibitem{EventHorizonTelescope:2019ths}
K.~Akiyama \textit{et al.} [Event Horizon Telescope],
Astrophys. J. Lett. \textbf{875}, no.1, L4 (2019)
[arXiv:1906.11241 [astro-ph.GA]].

\bibitem{EventHorizonTelescope:2019uob}
K.~Akiyama \textit{et al.} [Event Horizon Telescope],
Astrophys. J. Lett. \textbf{875}, no.1, L2 (2019)
[arXiv:1906.11239 [astro-ph.IM]].

\bibitem{EventHorizonTelescope:2022exc}
K.~Akiyama \textit{et al.} [Event Horizon Telescope],
Astrophys. J. Lett. \textbf{930}, no.2, L15 (2022)
[arXiv:2311.08697 [astro-ph.HE]].

\bibitem{EventHorizonTelescope:2022urf}
K.~Akiyama \textit{et al.} [Event Horizon Telescope],
Astrophys. J. Lett. \textbf{930}, no.2, L16 (2022)
[arXiv:2311.09478 [astro-ph.HE]].

\bibitem{EventHorizonTelescope:2022apq}
K.~Akiyama \textit{et al.} [Event Horizon Telescope],
Astrophys. J. Lett. \textbf{930}, no.2, L13 (2022)
[arXiv:2311.08679 [astro-ph.HE]].

\bibitem{EventHorizonTelescope:2022wok}
K.~Akiyama \textit{et al.} [Event Horizon Telescope],
Astrophys. J. Lett. \textbf{930}, no.2, L14 (2022)
[arXiv:2311.09479 [astro-ph.HE]].

\bibitem{EventHorizonTelescope:2022wkp}
K.~Akiyama \textit{et al.} [Event Horizon Telescope],
Astrophys. J. Lett. \textbf{930}, no.2, L12 (2022)
[arXiv:2311.08680 [astro-ph.HE]].

\bibitem{EventHorizonTelescope:2022xqj}
K.~Akiyama \textit{et al.} [Event Horizon Telescope],
Astrophys. J. Lett. \textbf{930}, no.2, L17 (2022)
[arXiv:2311.09484 [astro-ph.HE]].

\bibitem{Synge:1966okc}
J.~L.~Synge, 
\emph{Mon. Not. R. Astron. Soc.} \textbf{131}, 463--466 (1966).

\bibitem{Bardeen:1973tla}
J.~M.~Bardeen, 
\emph{Les Astres Occlus}, edited by C.~DeWitt and B.~S.~DeWitt (Gordon and Breach, New York, 1973), pp.~215--240.

\bibitem{Luminet:1979nyg}
J.~P.~Luminet, 
\emph{Astron. Astrophys.} \textbf{75}, 228--235 (1979).

\bibitem{Cunningham:1973tf}
R.~A.~Cunningham, E.~Gabathuler, M.~Sproston, and D.~G.~Taylor, 
\emph{Conf. Proc.} C \textbf{730508}, 298 (1973).

\bibitem{Bozza:2010xqn}
V.~Bozza,
Gen. Rel. Grav. \textbf{42}, 2269-2300 (2010)
[arXiv:0911.2187 [gr-qc]].

\bibitem{Falcke:1999pj}
H.~Falcke, F.~Melia and E.~Agol,
Astrophys. J. Lett. \textbf{528}, L13 (2000)
[arXiv:astro-ph/9912263 [astro-ph]].

\bibitem{Vries2000TheAS}
A.~de~Vries,
\emph{Class. Quantum Grav.} \textbf{17}, 123--144 (2000).

\bibitem{Shen:2005cw}
Z.~Q.~Shen, K.~Y.~Lo, M.~C.~Liang, P.~T.~P.~Ho and J.~H.~Zhao,
Nature \textbf{438}, 62 (2005)
[arXiv:astro-ph/0512515 [astro-ph]].

\bibitem{Yumoto:2012kz}
A.~Yumoto, D.~Nitta, T.~Chiba and N.~Sugiyama,
Phys. Rev. D \textbf{86}, 103001 (2012)
[arXiv:1208.0635 [gr-qc]].

\bibitem{Atamurotov:2013sca}
F.~Atamurotov, A.~Abdujabbarov and B.~Ahmedov,
Phys. Rev. D \textbf{88}, no.6, 064004 (2013)



\bibitem{Abdujabbarov:2015xqa}
A.~A.~Abdujabbarov, L.~Rezzolla and B.~J.~Ahmedov,
Mon. Not. Roy. Astron. Soc. \textbf{454}, no.3, 2423-2435 (2015)
[arXiv:1503.09054 [gr-qc]].


\bibitem{Cunha:2018acu}
P.~V.~P.~Cunha and C.~A.~R.~Herdeiro,
Gen. Rel. Grav. \textbf{50}, no.4, 42 (2018)
[arXiv:1801.00860 [gr-qc]].

\bibitem{Kumar:2018ple}
R.~Kumar and S.~G.~Ghosh,
Astrophys. J. \textbf{892}, 78 (2020)
doi:10.3847/1538-4357/ab77b0
[arXiv:1811.01260 [gr-qc]].

\bibitem{Afrin:2021ggx}
M.~Afrin and S.~G.~Ghosh,
Universe \textbf{8}, no.1, 52 (2022)
[arXiv:2112.15038 [gr-qc]].

\bibitem{Hioki:2009na}
K.~Hioki and K.~i.~Maeda,
Phys. Rev. D \textbf{80}, 024042 (2009)
[arXiv:0904.3575 [astro-ph.HE]].

\bibitem{Chen:2023wzv}
S.~Chen and J.~Jing,
JCAP \textbf{05}, 023 (2024)
[arXiv:2310.06490 [gr-qc]].

\bibitem{Li:2024abk}
X.~Q.~Li, H.~P.~Yan, X.~J.~Yue, S.~W.~Zhou and Q.~Xu,
JCAP \textbf{05}, 048 (2024)
[arXiv:2401.18066 [gr-qc]].

\bibitem{Amarilla:2010zq}
L.~Amarilla, E.~F.~Eiroa and G.~Giribet,
Phys. Rev. D \textbf{81}, 124045 (2010)
[arXiv:1005.0607 [gr-qc]].

\bibitem{Amarilla:2011fx}
L.~Amarilla and E.~F.~Eiroa,
Phys. Rev. D \textbf{85}, 064019 (2012)
[arXiv:1112.6349 [gr-qc]].

\bibitem{Amarilla:2013sj}
L.~Amarilla and E.~F.~Eiroa,
Phys. Rev. D \textbf{87}, no.4, 044057 (2013)
[arXiv:1301.0532 [gr-qc]].

\bibitem{Amir:2017slq}
M.~Amir, B.~P.~Singh and S.~G.~Ghosh,
Eur. Phys. J. C \textbf{78}, no.5, 399 (2018)
[arXiv:1707.09521 [gr-qc]].

\bibitem{Singh:2017vfr}
B.~P.~Singh and S.~G.~Ghosh,
Annals Phys. \textbf{395}, 127-137 (2018)
[arXiv:1707.07125 [gr-qc]].

\bibitem{Mizuno:2018lxz}
Y.~Mizuno, Z.~Younsi, C.~M.~Fromm, O.~Porth, M.~De Laurentis, H.~Olivares, H.~Falcke, M.~Kramer and L.~Rezzolla,
Nature Astron. \textbf{2}, no.7, 585-590 (2018)
[arXiv:1804.05812 [astro-ph.GA]].

\bibitem{Allahyari:2019jqz}
A.~Allahyari, M.~Khodadi, S.~Vagnozzi and D.~F.~Mota,
JCAP \textbf{02}, 003 (2020)
[arXiv:1912.08231 [gr-qc]].

\bibitem{Papnoi:2014aaa}
U.~Papnoi, F.~Atamurotov, S.~G.~Ghosh and B.~Ahmedov,
Phys. Rev. D \textbf{90}, no.2, 024073 (2014)
[arXiv:1407.0834 [gr-qc]].

\bibitem{Kumar:2020hgm}
R.~Kumar, S.~G.~Ghosh and A.~Wang,
Phys. Rev. D \textbf{101}, no.10, 104001 (2020)
[arXiv:2001.00460 [gr-qc]].

\bibitem{Kumar:2020owy}
R.~Kumar and S.~G.~Ghosh,
JCAP \textbf{07}, 053 (2020)
[arXiv:2003.08927 [gr-qc]].

\bibitem{Kumar:2019ohr}
R.~Kumar, B.~P.~Singh and S.~G.~Ghosh,
Annals Phys. \textbf{420}, 168252 (2020)
[arXiv:1904.07652 [gr-qc]].

\bibitem{Kumar:2017tdw}
R.~Kumar, B.~P.~Singh, M.~S.~Ali and S.~G.~Ghosh,
Phys. Dark Univ. \textbf{34}, 100881 (2021)
[arXiv:1712.09793 [gr-qc]].



\bibitem{Singh:2017xle}
B.~P.~Singh,
Annals Phys. \textbf{441}, 168892 (2022)
[arXiv:1711.02898 [gr-qc]].

\bibitem{Kumar:2020yem}
R.~Kumar, A.~Kumar and S.~G.~Ghosh,
Astrophys. J. \textbf{896}, no.1, 89 (2020)
[arXiv:2006.09869 [gr-qc]].

\bibitem{Kumar:2020sag}
R.~Kumar, S.~U.~Islam and S.~G.~Ghosh,
Eur. Phys. J. C \textbf{80}, no.12, 1128 (2020)
[arXiv:2004.12970 [gr-qc]].

\bibitem{Kumar:2020ltt}
R.~Kumar and S.~G.~Ghosh,
Class. Quant. Grav. \textbf{38}, no.8, 8 (2021)
[arXiv:2004.07501 [gr-qc]].

\bibitem{Kumar:2019pjp}
R.~Kumar, S.~G.~Ghosh and A.~Wang,
Phys. Rev. D \textbf{100}, no.12, 124024 (2019)
[arXiv:1912.05154 [gr-qc]].

\bibitem{Kumar:2024cnh}
A.~Kumar, D.~V.~Singh and S.~Upadhyay,
Int. J. Mod. Phys. A \textbf{39}, no.31, 2450136 (2024)
[arXiv:2412.14230 [gr-qc]].

\bibitem{Vishvakarma:2023csw}
B.~K.~Vishvakarma, D.~V.~Singh and S.~Siwach,
Eur. Phys. J. Plus \textbf{138}, no.6, 536 (2023)
[arXiv:2304.14754 [gr-qc]].

\bibitem{Ghosh:2020spb}
S.~G.~Ghosh, R.~Kumar and S.~U.~Islam,
JCAP \textbf{03}, 056 (2021)
[arXiv:2011.08023 [gr-qc]].

\bibitem{Guo:2020zmf}
M.~Guo and P.~C.~Li,
Eur. Phys. J. C \textbf{80}, no.6, 588 (2020)
[arXiv:2003.02523 [gr-qc]].

\bibitem{Afrin:2021wlj}
M.~Afrin and S.~G.~Ghosh,
Astrophys. J. \textbf{932}, no.1, 51 (2022)
[arXiv:2110.05258 [gr-qc]].

\bibitem{Vagnozzi:2022moj}
S.~Vagnozzi, R.~Roy, Y.~D.~Tsai, L.~Visinelli, M.~Afrin, A.~Allahyari, P.~Bambhaniya, D.~Dey, S.~G.~Ghosh and P.~S.~Joshi, \textit{et al.}
Class. Quant. Grav. \textbf{40}, no.16, 165007 (2023)
[arXiv:2205.07787 [gr-qc]].

\bibitem{Vagnozzi:2019apd}
S.~Vagnozzi and L.~Visinelli,
Phys. Rev. D \textbf{100}, no.2, 024020 (2019)
[arXiv:1905.12421 [gr-qc]].

\bibitem{Afrin:2021imp}
M.~Afrin, R.~Kumar and S.~G.~Ghosh,
Mon. Not. Roy. Astron. Soc. \textbf{504}, no.4, 5927-5940 (2021)
[arXiv:2103.11417 [gr-qc]].

\bibitem{Gao:2023mjb}
X.~J.~Gao, T.~T.~Sui, X.~X.~Zeng, Y.~S.~An and Y.~P.~Hu,
Eur. Phys. J. C \textbf{83}, 1052 (2023)
[arXiv:2311.11780 [gr-qc]].

\bibitem{Ghosh:2022jfi}
S.~G.~Ghosh and R.~K.~Walia,
doi:10.1142/9789811269776\_0084
[arXiv:2203.07775 [gr-qc]].

\bibitem{Li:2022eue}
S.~Li, T.~Mirzaev, A.~A.~Abdujabbarov, D.~Malafarina, B.~Ahmedov and W.~B.~Han,
Phys. Rev. D \textbf{106}, no.8, 084041 (2022)
[arXiv:2207.10933 [gr-qc]].

\bibitem{Sengo:2022jif}
I.~Sengo, P.~Cunha, V.P., C.~A.~R.~Herdeiro and E.~Radu,
JCAP \textbf{01}, 047 (2023)
[arXiv:2209.06237 [gr-qc]].

\bibitem{Liu:2024lbi}
W.~Liu, D.~Wu, X.~Fang, J.~Jing and J.~Wang,
JCAP \textbf{08}, 035 (2024)
[arXiv:2406.00579 [gr-qc]].

\bibitem{Umarov:2025btg}
D.~Umarov, O.~Yunusov, F.~Atamurotov, A.~Abdujabbarov and S.~G.~Ghosh,
Chin. Phys. C \textbf{49}, no.5, 055102 (2025)

\bibitem{Ahmed:2025zdc}
F.~Ahmed, S.~U.~Islam and S.~G.~Ghosh,
JHEAp \textbf{46}, 100350 (2025)

\bibitem{Turakhonov:2025ojy}
Z.~Turakhonov, F.~Atamurotov, S.~G.~Ghosh and A.~Abdujabbarov,
Phys. Dark Univ. \textbf{48}, 101880 (2025)

\bibitem{Singh:2023zmy}
B.~P.~Singh,
Phys. Dark Univ. \textbf{42}, 101279 (2023)
[arXiv:2301.00956 [gr-qc]].

\bibitem{Singh:2022dqs}
B.~P.~Singh, M.~S.~Ali and S.~G.~Ghosh,
Commun. Theor. Phys. \textbf{77}, no.7, 075407 (2025)
[arXiv:2207.11907 [gr-qc]].

\bibitem{KumarWalia:2022aop}
R.~Kumar Walia, S.~G.~Ghosh and S.~D.~Maharaj,
Astrophys. J. \textbf{939}, no.2, 77 (2022)
[arXiv:2207.00078 [gr-qc]].

\bibitem{Liu:2020ola}
C.~Liu, T.~Zhu, Q.~Wu, K.~Jusufi, M.~Jamil, M.~Azreg-A\"\i{}nou and A.~Wang,
Phys. Rev. D \textbf{101}, no.8, 084001 (2020)
[erratum: Phys. Rev. D \textbf{103}, no.8, 089902 (2021)]
[arXiv:2003.00477 [gr-qc]].

\bibitem{Brahma:2020eos}
S.~Brahma, C.~Y.~Chen and D.~h.~Yeom,
Phys. Rev. Lett. \textbf{126}, no.18, 181301 (2021)
[arXiv:2012.08785 [gr-qc]].

\bibitem{KumarWalia:2022ddq}
R.~Kumar Walia,
JCAP \textbf{03}, 029 (2023)
[arXiv:2207.02106 [gr-qc]].

\bibitem{Islam:2022wck}
S.~U.~Islam, J.~Kumar, R.~Kumar Walia and S.~G.~Ghosh,
Astrophys. J. \textbf{943}, no.1, 22 (2023)
[arXiv:2211.06653 [gr-qc]].


\bibitem{Afrin:2022ztr}
M.~Afrin, S.~Vagnozzi and S.~G.~Ghosh,
Astrophys. J. \textbf{944}, no.2, 149 (2023)
[arXiv:2209.12584 [gr-qc]].

\bibitem{Yang:2022btw}
J.~Yang, C.~Zhang and Y.~Ma,
Eur. Phys. J. C \textbf{83}, no.7, 619 (2023)
[arXiv:2211.04263 [gr-qc]].

\bibitem{Singh:2023ops}
B.~P.~Singh, R.~Kumar and S.~G.~Ghosh,
New Astron. \textbf{99}, 101945 (2023)

\bibitem{Hou:2021okc}
Y.~Hou, M.~Guo and B.~Chen,
Phys. Rev. D \textbf{104}, no.2, 024001 (2021)
[arXiv:2103.04369 [gr-qc]].

\bibitem{Vagnozzi:2022tba}
S.~Vagnozzi and L.~Visinelli,
Res. Notes AAS \textbf{6}, no.5, 106 (2022)
[arXiv:2205.11314 [astro-ph.GA]].

\bibitem{Banerjee:2022jog}
I.~Banerjee, S.~Chakraborty and S.~SenGupta,
Phys. Rev. D \textbf{106}, no.8, 084051 (2022)
[arXiv:2207.09003 [gr-qc]].

\bibitem{Lemos:2024wwi}
A.~S.~Lemos, J.~A.~V.~Campos and F.~A.~Brito,
Phys. Rev. D \textbf{110}, no.6, 064079 (2024)
[arXiv:2407.04609 [gr-qc]].

\bibitem{Singh:2024rnh}
B.~P.~Singh,
Annals Phys. \textbf{470}, 169803 (2024)
[arXiv:2409.07951 [gr-qc]].


\bibitem{Chandrasekhar:1985kt}
S.~Chandrasekhar, 
\emph{The Mathematical Theory of Black Holes} 
(Oxford University Press, New York, 1985).

\bibitem{Carter:1971zc}
B.~Carter,
Phys. Rev. Lett. \textbf{26}, 331-333 (1971)

\bibitem{Frolov:2014dta}
A.~V.~Frolov and V.~P.~Frolov,
Phys. Rev. D \textbf{90}, no.12, 124010 (2014)
[arXiv:1408.6316 [gr-qc]].

\bibitem{bardeen1973}
J.~Bardeen, ``Black holes,'' in \emph{Black Holes}, edited by C.~DeWitt and B.~S.~DeWitt (Gordon and Breach, New York, 1973).

\bibitem{Bambi:2019tjh}
C.~Bambi, K.~Freese, S.~Vagnozzi and L.~Visinelli,
Phys. Rev. D \textbf{100}, no.4, 044057 (2019)
[arXiv:1904.12983 [gr-qc]].

\end{thebibliography}
\end{document}